\documentclass[%
 aip,
 apl,
 floatfix,
 amsmath,amssymb,
 reprint,%
]{revtex4-1}

\usepackage{graphicx}
\usepackage{dcolumn}
\usepackage{bm}

\usepackage[utf8]{inputenc}
\usepackage[T1]{fontenc}
\usepackage{dsfont}
\usepackage{mathptmx}
\usepackage{etoolbox}
\usepackage{siunitx}
\usepackage{listings}
\usepackage{subcaption}
\usepackage[dvipsnames]{xcolor}
\usepackage{hyperref}
\usepackage{booktabs}

\makeatletter
\def\@email#1#2{%
 \endgroup
 \patchcmd{\titleblock@produce}
  {\frontmatter@RRAPformat}
  {\frontmatter@RRAPformat{\produce@RRAP{*#1\href{mailto:#2}{#2}}}\frontmatter@RRAPformat}
  {}{}
}%
\makeatother
\begin{document}

\preprint{AIP/123-QED}

\title[Gradient-based optimization of scatterer
arrangements based on the T-matrix
method]{Gradient-based optimization of scatterer
arrangements based on the T-matrix
method}
\author{N. Asadova}
 \email{nigar.asadova@kit.edu}
 \author{J. D. Fischbach}%
\affiliation{ 
 Institute of Nanotechnology, Karlsruhe Institute of Technology (KIT), Karlsruhe, Germany
}%
\author{R. Vallée}%
\affiliation{ 
Centre de Recherche Paul Pascal, University of Bordeaux, Pessac, France
}%

\author{O. Kuster}
\affiliation{Institute of Theoretical Solid State Physics, Karlsruhe Institute of Technology (KIT), Karlsruhe, Germany}

\author{Y. Augenstein}%
\affiliation{ 
Flexcompute Inc, Belmont, MA, USA
}%

\author{D. Vovchuk}
\affiliation{Institute of Photonics, Electronics and Telecommunications, Riga Technical University, Riga, Latvia}
\affiliation{School of Electrical Engineering, Tel Aviv University, Tel Aviv, Israel}

\author{A. Kharchevskii}
\affiliation{School of Electrical Engineering, Tel Aviv University, Tel Aviv, Israel}

\author{P. Ginzburg}
\affiliation{School of Electrical Engineering, Tel Aviv University, Tel Aviv, Israel}

\author{C. Rockstuhl}
 \affiliation{ 
Institute of Nanotechnology, Karlsruhe Institute of Technology (KIT),  Karlsruhe, Germany
}%

 \affiliation{ 
 Institute of Theoretical Solid State Physics, Karlsruhe Institute of Technology (KIT), Karlsruhe, Germany
}%
\affiliation{Center for Integrated Quantum Science and Technology (IQST), Karlsruhe Institute of Technology,  Karlsruhe, Germany}

\begin{abstract}
The demand for inverse design is increasing as the ability to fabricate sub-10 nm features expands the design space significantly. Efficient inverse design benefits from differentiable models of light-matter interaction. While traditional full-wave solvers based on finite differences, finite elements, or Fourier modal methods have already been presented for that purpose, a dedicated tool adapted for performing multiple scattering simulations is still lacking. To overcome this limitation, we provide a multiple-scattering framework compatible to automatic differentiation, suitable for treating periodic and non-periodic arrangements of scatterers. It yields exact gradients regarding geometric and positional parameters in finite clusters and infinite metasurfaces. In this work, we use spheres as the building blocks to demonstrate the framework's capabilities as a standalone tool. However, the framework is adaptable to arbitrarily shaped scatterers, provided the individual T-matrices are calculated using differentiable Maxwell solvers. Since the gradients are obtained simultaneously in a single backward pass, the framework is well-suited for moderately dimensional problems. It is also possible to combine multiple performance goals into a single objective function. The versatility of our method is illustrated in proof-of-concept examples that focus on Kerker-type physics. In the first example, a finite cluster of scatterers is optimized in order to reach a high forward-to-backward scattering ratio, and we show experimental feasibility of the designs. In the second example, a metasurface made from multiple scatterers per unit cell is designed to maximize the reflectance contrast between orthogonal linear polarizations of the incident light. We make the framework publicly available at \url{https://github.com/tfp-photonics/dreams}.

\end{abstract}

\maketitle

\section{Introduction}
With recent advances in the fabrication of nanophotonic devices, numerous design parameters can be tailored to achieve a desired optical response. This increase in degrees of freedom necessitates solving the inverse problem efficiently. In other words, we need to identify optimal design parameters so the nanophotonic device exhibits the desired optical response.

The most basic approach is to perform a parameter sweep and select the best performance among the calculated objective values. However, this quickly becomes infeasible with an increasing number of parameters, a phenomenon known as the curse of dimensionality. Therefore, over the decades, a plethora of more advanced approaches were introduced~\cite{molesky2018inverse, campbell2019review, yao2019intelligent, wang2022advancing, bennet2024illustrated}. The problem to be solved within the framework of inverse design is to adjust each degree of freedom to reach the optimum for a desired objective function. Inverse design employs various strategies such as deep learning~\cite{wiecha2021deep, ma2022benchmarking}, topology optimization~\cite{sell2017large, lin2019topology, stich2025inverse}, metaheuristic algorithms~\cite{mirzaei2015all, li2018broadband, wiecha2019design}, or probabilistic algorithms~\cite{schneider2017global, wray2024optical}. 

However, scalability remains a critical challenge~\cite{eggensperger2013towards, angeris2021heuristic}. As the number of design parameters increases, the computational time required by heuristic methods often increases exponentially, making them less suitable for high-dimensional design problems~\cite{campbell2019review}. Among different approaches, deep learning models provide practically immediate results after the training phase is completed. However, training depends on the availability of an extensive dataset. The data generation process scales exponentially with the dimensionality of the design space~\cite{jiang2021deep}, and in the majority of cases, each particular problem requires a dedicated training set. This renders such approaches computationally expensive in scenarios where the training data must be explicitly generated solely to train a neural network to solve a specific inverse design problem. 

Therefore, in high-dimensional parameter spaces, gradient-based optimization methods are the primary choice for solving inverse problems. The known gradients provide the shortest path to a local optimum. Although the optimum is not global, the curse of dimensionality can become a blessing~\cite{li2022empowering}. For very high-dimensional parameter spaces, the landscape is expected to flatten out such that the local maximum or minimum is close in value to the global one. However, this effect depends on the specific nature of the problem and must be empirically tested. Beyond the algorithmic considerations, it is also useful to understand the fundamental limits on what is achievable for a given material and design region~\cite{molesky2022t, chao2022physical}.

The calculation of gradients for complex computational code can be implemented in different ways. Numerical implementation with finite differences is typically not considered, since it requires performing one additional simulation for each parameter variation. In the adjoint method, a second problem has to be manually formulated and solved, and the solution provides the gradient of the objective function. It is usually formulated for PDE problems. The adjoint method is extensively used in density-based topology optimization, where each pixel or voxel in the device elements serves as a degree of freedom~\cite{bendsoe2013topology}. This easily amounts to a total of $10^{3}$-$10^{9}$ degrees of freedom. The resulting optimized shapes are free-form, and potential fabrication difficulties should be handled as well~\cite{piggott2017fabrication, augenstein2020inverse, hammond2021photonic}. 
 
An alternative approach to obtaining gradients is automatic differentiation (AD) -- an automated approach that evaluates exact derivatives of a composition of elementary functions. As it is possible to spend the same amount of time to calculate the derivative with respect to all parameters, regardless of the number of parameters, a single optimization step takes roughly twice the time of a single simulation. Hybrid approaches replace manual derivation of the adjoint equation with AD~\cite{luce2023merging}. However, fully differentiable implementations go further, re-implementing the solver entirely within a framework with AD capabilities for ease of use and high performance.  

Formulating tools for scientific computations in an automatically differentiable manner requires dedicated effort and careful implementation. In essence, every elementary function needs to be formulated as differentiable, something which poses challenges. Nevertheless, with the appreciation that gradient information is key to inverse design in high-dimensional parameter spaces, multiple efforts have been dedicated to formulating various methods that solve Maxwell's equations in an automatically differentiable manner. These include general-purpose full-wave Maxwell solvers, such as the finite-difference time-domain (FDTD) method~\cite{hughes2019forward, hooten2025automatic}, the finite element method (FEM)~\cite{vial2022open, xue2023jax}, and others.  

A special-purpose computational method that was already formulated in an automatically differentiable manner is the rigorous coupled-wave analysis (RCWA) method, also known as the Fourier modal method (FMM)~\cite{liu2012s4, hugonin2021reticolo}. It has been extensively used within the AD framework~\cite{colburn2021inverse, zhu2020differentiable, so2023multicolor, kim2023torcwa, jin2020inverse} or adjoint method~\cite{backer2019computational}, and can be combined with topology optimization~\cite{lin2019topology, phan2019high}.  The discrete dipole approximation has also been used with an adjoint formulation for the optimization of multiple-scattering systems. Another semi-analytical approach with a recently developed differentiable implementation is the hybrid coupled dipole method~\cite{ponomareva2025torchgdm}. However, each of these automatically differentiable formulations inherits the pros and cons of the underlying computational method. Therefore, having a wide range of tools available that can be flexibly applied to specific problems is highly beneficial for further developing nanophotonics.

This contribution focuses on the automatic differentiation of a computational framework for nanophotonic scattering problems based on the T-matrix method. The benefit of the method lies in its semi-analytical treatment of the light-matter interaction with the elementary building blocks expressed using the T-matrix. The T-matrix captures how an incident field, expanded into vector spherical waves, is converted into a scattered field, which is also expanded into vector spherical waves. Once the T-matrix of an object is known, complex arrangements of many scatterers can be studied highly efficiently based on an underlying algebraic formulation. The scatterers can be arranged randomly or in a periodic pattern, forming metasurfaces or metamaterials. Such versatility renders the T-matrix applicable to a wide range of nanophotonic problems~\cite{yannopapas2011layer, pal2025dark, tanaka2020chiral, salerno2022loss}. Our implementation follows the open-source \textit{treams} package~\cite{beutel2024treams} and integrates AD capabilities into the multiscattering code. 

A central milestone of our work is the development of a differentiable computational framework for calculating the optical response of complex arrangements of scatterers. The framework that we develop here enables differentiation with respect to input parameters, including not only the radii of individual spheres but also the positions of spheres or arbitrarily shaped scatterers in arrangements such as periodic arrays with complex unit cells containing multiple scatterers. Here, we use the term ``moderately dimensional`` to refer to optimization problems with approximately up to a hundred scatterers, with multipole truncation orders in the range $l_{\max}=3-7$, which is sufficient for most of the nanophotonics problems. Depending on the parameterization, this corresponds to in the order of hundreds of optimizable variables.

Earlier implementations used analytically calculated derivatives of Mie coefficients, and the final derivatives were obtained using the adjoint method~\cite{zhan2018inverse, zhan2019controlling}. 
These works did not utilize the AD framework, and the parametrization of scatterer positions was not integrated, as it requires more manual derivations. Recent publications applied gradient-free optimization to infinitely extended core-shell cylinders~\cite{igoshin2025inverse} and the arrangement of spheres~\cite{tsukerman2025diffusion}. Gradient-based optimization for chains of infinite cylinders was performed with Gaussian parameterization of positions~\cite{fischbach2025framework}. 

The article is structured as follows. In Section~\ref{method}, the multiscattering formalism is introduced, and the fundamentals of AD are outlined. In Section~\ref{results}, an arrangement of spheres is optimized to achieve the maximum scattering ratio between the forward and backward hemispheres. Next, the optimization results for an example of a metasurface with a complex unit cell composed of spheres, optimized for polarization-selective reflectance, are discussed. By making the underlying Python source codes developed in this work publicly available on GitHub (\url{https://github.com/tfp-photonics/dreams}), we expect a substantial contribution to the further development of the field of nanophotonics that relies on scattering structures.

\section{Automatic differentiation of multiscattering framework}\label{method}
\subsection{Multiscattering formalism}

The T-matrix formalism is a versatile approach that simplifies the computation of the optical response for various arrangements of scatterers. Here, the T-matrix represents the linear relationship between expansion coefficients of the regular vector spherical waves that expand the incident field and the singular vector spherical waves that expand the scattered field:

\begin{equation}
\mathbf{p}=\mathbf{T} \, \mathbf{a}
\, .
\end{equation}
$\mathbf{p}$ and $\mathbf{a}$ are vectors containing the expansion coefficients for the scattered field and the incident field, respectively:
\begin{align}
  \mathbf{E}_{\text{inc}}(\mathbf{r}, \omega) &= \sum_{l=1}^{\infty}\sum_{m=-l}^{l}\left[a_{lm}^\mathrm{e}(\omega) \mathbf{N}^{(1)}_{l m }(\mathbf{r}, \omega) + a_{lm}^\mathrm{m}(\omega) \mathbf{M}^{(1)}_{l m }(\mathbf{r}, \omega)\right]
  \\
    \mathbf{E}_{\text{sca}}(\mathbf{r}, \omega) &= \sum_{l=1}^{\infty}\sum_{m=-l}^{l}\left[ p_{lm}^\mathrm{e} (\omega) \mathbf{N}^{(3)}_{l m }(\mathbf{r}, \omega) + p_{lm}^\mathrm{m}(\omega) \mathbf{M}^{(3)}_{l m }(\mathbf{r}, \omega)\right]\, .
\end{align}
The fields are expanded in transverse electric (TE) and transverse magnetic (TM) vector spherical waves $\mathbf{M}_{l m }(\mathbf{r}, \omega)$ and $\mathbf{N}_{l m }(\mathbf{r}, \omega)$. The superscripts correspond to regular (1) and singular (3) fields~\cite{jackson1998classical}. The T-matrix $\mathbf{T}$ represents the most comprehensive information on the linear optical response of a scatterer and can be obtained semi-analytically for basic objects, i.e., spheres and infinite cylinders, or numerically for arbitrarily shaped scatterers.
The number of multipolar orders in the expansion can be truncated at a specific value, beyond which the contribution to the optical response can be considered negligible. This approach, therefore, offers clear advantages over the commonly used dipole approximation.

For arrangements of scatterers, a modified expression holds that accounts for the contribution of the scattered field coming from other scatterers to the incident field on each scatterer:
\begin{equation}\label{toglobal}
 \mathbf{p}= (\mathds {1} - \mathbf{T} \mathbf{C}^{(3)})^{-1} \mathbf{T} \mathbf{a}
\, , 
\end{equation}
where $\mathbf{T}$ is a block-diagonal matrix, where each scatterer response is defined in its local coordinate system. All the translation coefficients are packed in the matrix $\mathbf{C}^{(3)}$, and $\mathbf{C}_{ij}^{(3)}$ depends on the distance from the $j$-th scatterer to the $i$-th scatterer. Please note that this is what we refer to as a local description, as each scatterer is represented by its T-matrix.

We can also employ a global T-matrix formulation, where the entire scattering response of an ensemble of scatterers is expressed relative to a specific origin~\cite{cruzan1962translational, suryadharma2017studying}. This is beneficial since the T-matrix shrinks in size for problems with closely arranged scatterers. 
For the scattering problem of a periodic lattice, $\mathbf{C}_{ij}^{(3)}$ includes a sum over an infinite number of lattice sites. The direct evaluation of the sum converges poorly, so the Ewald method is employed~\cite{ewald1921berechnung, beutel2023unified}. By expanding the resulting T-matrix from the periodic spherical wave basis into the plane wave basis, the S-matrix of a single layer can be obtained that relates the incoming and outgoing plane waves. Finally, the single infinite layer can be stacked with another two-dimensional lattice, an interface, or a homogeneous slab. Thus, the complete framework can simulate a diverse set of scattering arrangements.   

\subsection{Automatic differentiation}

Now, with this framework, we can express the optical response of various types of photonic materials composed of an ensemble of isolated scatterers. But for the inverse problem, we need to calculate the gradients, i.e., the change of an objective function derived from a forward simulation, depending on any of the degrees of freedom that we consider. To do so, we will automatically differentiate the code that solves the forward problem.    

The basic idea of AD is to apply the chain rule to numerical values and not symbolic expressions. Consider $f$ : $\mathbb{R}^m \rightarrow \mathbb{R}^n$ having $m$ input values and $n$ output values as a sequence of elementary operations. In an optimization task, the derivative of each output with respect to each input variable is required, which makes up a Jacobian of the size $n \times m$. Each intermediate variable in the calculation can be associated with a derivative. Exact derivatives are known for each primitive function. They are linked according to the chain rule, so that the derivative of the function output with respect to an input variable can be easily traced. 


 In this work, we focus on reverse-mode automatic differentiation, suitable for our typical optimization setting.
For many real-world optimization tasks, the output is typically a single value, while the input comprises a considerably larger number of variables. Therefore, it is more advantageous to compute a row of the Jacobian at a time, which includes derivatives of a single output value with respect to all of the inputs. The vector Jacobian product (VJP) is defined as:
\begin{equation}
  \text{VJP}\left(\mathbf{f}(\mathbf{v}, \mathbf{x})\right)=\mathbf{v}^T \frac{\partial{\mathbf{f}}}{\partial{\mathbf{x}}}(\mathbf{x})\, ,
\end{equation}
where $\frac{\partial{\mathbf{f}}}{\partial{\mathbf{x}}}$ is the Jacobian specified above, and $\mathbf{v} (\mathbf{x})$ is a vector of incoming variables.
Here, one should keep in mind that intermediate results have to be stored before starting the backward pass to evaluate the gradients. 
Reverse mode AD is similar to the adjoint method mentioned earlier, while the latter does not assume a fully automatic process, rather a derivation and embedding of the adjoint equations is required. However, even for the simplified automatic differentiation process, there are numerous pitfalls to be aware of~\cite{huckelheim2024taxonomy}. For a detailed overview of differentiable programming, we refer the reader to a recent review~\cite{blondel2024elements}.

JAX is used in the current version of the code to perform automatic differentiation~\cite{jax2018github}. This approach is a combination of operator overloading, similar to PyTorch,~\cite{paszke2019pytorch}, and source code transformation, which is the approach in TensorFlow ~\cite{abadi2016tensorflow}. JAX also provides an option to define derivatives for custom primitive functions, which was used to integrate special functions required in the T-matrix approach into the computational flow. For example, spherical Bessel and Hankel functions are not defined in JAX, so they are introduced as primitive functions with their derivative rule manually defined. An example of a primitive function definition is provided in Appendix~\ref{appendix:derivation}. 

In contrast to the original package \textit{treams}, the core functions are modified because the underlying implementation in Cython is not supported directly in the JAX framework, and they have to be transformed into Python functions. We avoid Python loops, which can introduce computational bottlenecks, opting instead for vectorized operations where it is reasonable. The test functions were computed with the original package and the current code to ensure that the different implementations do not introduce numerical discrepancies. In the Supplementary Material (Sec.~S1) we verify the gradients computed with automatic differentiation by comparison to central finite differences over a range of step sizes. Generally, gradient checks can be performed conveniently using the \texttt{check\_grads} utility of JAX. Finally, while JAX uses single precision by default, we use double precision throughout optimizations to ensure higher numerical accuracy.

\section{Optimization results}\label{results}
The final computational tools are made publicly available~\cite{dreams}. To demonstrate the general applicability of our approach and ensure consistency, we present two examples from the field of Kerker-physics that highlight its strengths. These examples are intended to be illustrative, and the methodology can be applied to a wide range of other problems. We begin with an example where we inversely design a finite cluster of scatterers, in order to optimize its integrated forward-to-backward scattering ratio. It is followed by an example where a unit cell of a metasurface, consisting of multiple scatterers, is optimized, maximizing its reflectance for a specified polarization. 

To evaluate our gradient-based approach against a gradient–free optimization technique, we performed optimizations using Scalable Constrained Bayesian optimization and included the relevant information in the Supplementary Material (Sec.~S3) for the first example. Across the hyperparameter sweep and initial randomizations tested, we achieved at most half of the performance of our approach. In addition, we used the setup of the second example, and performed one gradient-based optimization iteration, this time using the FDTD solver Tidy3D and the FEM solver JCMsuite for the forward simulation and gradient computation, as described in the same section of the Supplementary Material. This comparison further demonstrates that the semi-analytical method indeed provides a computational speedup.

\subsection{Finite cluster}
If the electric and magnetic dipole responses of a scatterer have equal strength and are in phase at a given wavelength, zero dipolar backward scattering (first Kerker condition) is achieved. If they have opposite phase shift, near-zero dipolar forward scattering (second Kerker condition) takes place~\cite{alu2010does}.
According to the generalized Kerker effect \cite{alaee2015generalized, liu2018generalized}, higher-order multipoles can be leveraged to achieve zero total backscattering. This, in essence, requires the constructive interference of all scattering contributions from all multipolar orders in the forward direction. In contrast, they destructively interfere in the backward direction. Even more intricate patterns emerge when specific combinations of multipoles are engineered~\cite{miroshnichenko2015nonradiating, shamkhi2019transverse}.  Generally, an object with a scattering cross-section exceeding the single-channel regime is called a superscatterer~\cite{ruan2010superscattering, ziolkowski2017using, krasikov2021multipolar}. 

In this work, we consider scattering integrated over the entire forward and backward hemispheres rather than at a single direction. 
Hemispherical ($2\pi$) integration is shown to deliver major advantages over single-angle detection, especially for highly anisotropic nanostructures that scatter light into sharp forward or backward lobes.  Experimental studies demonstrate that relying on direct or limited-angle measurements can introduce errors of 10\% or more, often underestimating or misrepresenting total scattered power~\cite{lindstrom2000reflectance, payne2015broadband, kim2005development}. Only $2\pi$ integration supports accurate modeling and optimization, crucial for applications like light-trapping and absorption in solar cells~\cite{payne2015broadband, jung2011dyadic, ulriksen2019plasmon}.

For these reasons, the objective is defined as:  
\begin{equation}
    \mathrm{FOM} = \frac{\int_{S_f} \left| E_{\mathrm{sca}} \right| ^2 \,\mathrm{d}S}
    {\int_{S_b} \left| E_{\mathrm{sca}} \right| ^2 \,\mathrm{d}S},
\end{equation}
where $E_{\mathrm{sca}}$ denotes the scattered field, and $S_f$ and $S_b$ are the surfaces of the forward and backward hemispheres, respectively. The integrations are performed in the far field. The integrals are evaluated using a discrete quadrature method.
We maximize the F/B scattering ratio from a cluster of particles consisting of a finite number of spheres of a given material. The relative position of the spheres and their size are subject to optimization.
Computations employ a local basis for each T-matrix without expanding in a global basis. However, for a later discussion of the multipolar contributions to the scattering response of the entire cluster, we resort to a global basis. 

\begin{figure}[!htbp]
  \centering
  \includegraphics[width=\linewidth]{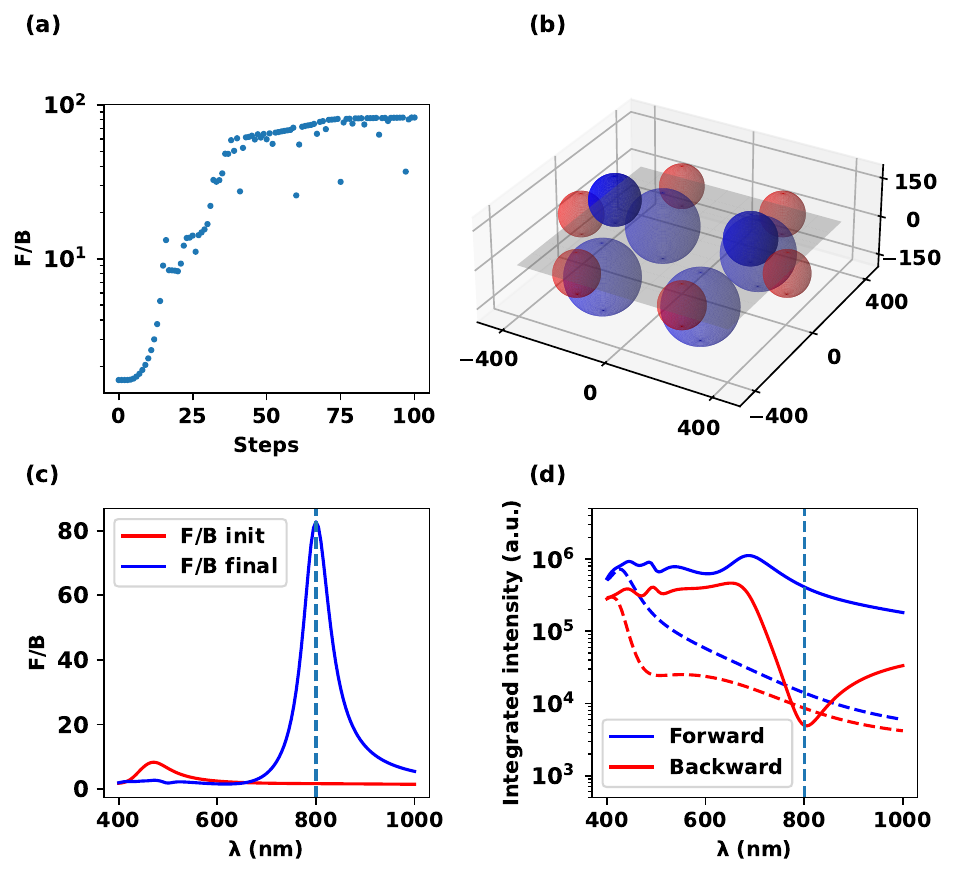}
  \caption{(a) Convergence of the F/B ratio during optimization at the design wavelength of 800~nm when optimizing the sample from a given initial configuration, as explained in the main text. (b)  3D view of the cluster, with the z = 0 plane shown in grey and spheres on the plane rendered with lower transparency. The initial configuration is indicated in red, and the optimized structure in blue. (c) F/B ratio spectrum shows an enhancement at the design wavelength. 
  (d) The scattered field intensity integrated over the forward (blue) and backward (red) hemispheres. Dashed curves indicate the initial arrangement and solid curves the optimized one.  The dashed vertical line marks the design wavelength. 
}
\label{sca1}\end{figure} 

To ensure that all optimization problems comply with the necessary constraints, we incorporate them using a nonlinear optimizer. We opt for the method of moving asymptotes (MMA)~\cite{svanberg2002class} as it handles both bounds and nonlinear constraints. The constraints must be differentiable and evaluated at each iteration. The algorithm builds a sequence of convex and separable approximations to both the objective and the nonlinear constraints. We impose the no-overlap condition of adjacent spheres as a nonlinear constraint on the positions of the spheres for both finite and infinite arrangements. This is implemented by evaluating for all sphere pairs the quantity
\[
r_i + r_j - d_{ij} +s,
\]
where \(r_i\) and \(r_j\) are the particle radii, \( d_{ij} = \|\mathbf{x}_i - \mathbf{x}_j\|_2 \)  is the Euclidean distance, and \(s \ge 0\) is an optional safety margin.
The constraint value is taken as the maximum of this quantity over all
pairs, and requiring it to remain non-positive guarantees non-overlap.  Besides enforcing physical non-overlap, this constraint also serves as a numerical safeguard by preventing configurations, for which the translation coupling terms and thus the corresponding multiscattering linear system given in Eq.~\ref{toglobal} may become poorly conditioned. In the implementation, the linear system is evaluated numerically using direct dense solve rather than by forming an inverse operator.

Additionally, bound constraints specify the maximum and minimum values for the design variables, ensuring that the objective is not evaluated outside permissible limits. Bounds are set on the minimum radii of the spheres to adhere to possible fabrication constraints. Another potential issue for very small spheres is the significance of quantum effects, which must be accounted for with appropriate approximations.  Consequently, we enforce a minimum radius of 5~nm. 

For a physically accurate simulation, the maximum radius of the spheres and the minimal distance between them must be consistent with the number of multipoles used. This is ensured by comparing the results against simulations using a higher number of multipoles. For these reasons, we do not restrict the radii to an upper limit.  In principle, it is possible to adapt the number of multipoles on the fly based on the current size parameters, but this option was not chosen here. Meanwhile, changing the number of spheres would modify the number of parameters. Therefore, it is not supported within a single run. Furthermore, since the local basis formalism is used, it is also not necessary to control whether the distance between spheres becomes too large, which would require more multipole orders in the global basis expansion than what was sufficient in the initial arrangement. With a local basis, we include up to octupole order throughout the optimizations.

The optimization results are presented in Fig.~\ref{sca1}. Spheres with a relative permittivity of 6.25 in vacuum are initially arranged along a circle with a radius of 400~nm. This arrangement is depicted with red spheres in Fig.~\ref{sca1}~(b). In the chosen example, we have opted to consider six spheres with a radius of 80~nm. Similar results were obtained with a larger or smaller number of spheres with different radii as initial conditions. However, if the number is too large, it overly constrains the optimization search space, and the performance degrades. The initial arrangement should be set on a larger circle to provide sufficient design freedom. Additional results for extended initial configurations, including cases with added outer rings, as well as the corresponding scaling of the end-to-end optimization time and the achieved normalized objective function, are provided in Section S2 of the Supplementary Material for completeness. 

 To set a baseline, we initially computed the F/B scattering ratio of this given structure upon illumination with a $y$-polarized plane wave propagating in the $z$-direction in the wavelength range from 400 to 1,000~nm. The F/B scattering ratio of this initial configuration is shown in Fig.~\ref{sca1}~(c). A clear peak of 10 is observed at around 450~nm, which can be attributed to the single particle magnetic dipole resonance.

Optimization of the F/B scattering ratio is performed at the design wavelength of 800~nm. The optimization variables are the 3D positions and radii of the spheres, which gives us $4N=24$ design parameters. The number of spheres $N$ and the multipole truncation $l_{\max}$ are fixed during an optimization run. The positions and radii of the spheres are adjusted in each iteration, respecting the overlap constraints. We initialize the spheres with small radii to avoid early overlap constraint violations during optimization. As can be seen in Fig.~\ref{sca1}~(c), the final F/B ratio reaches 80 at the design wavelength. The separately plotted numerator and denominator in Fig.~\ref{sca1}~(d) demonstrate that the stronger increase in the forward-scattering contribution is responsible for the increase of the total scattering. Figure~\ref{sca1}~(a) shows that the algorithm has converged after 40 iterations, where the achievable F/B scattering ratio reaches a plateau. 
\begin{figure}[ht]
  \centering
  \includegraphics[width=\linewidth]{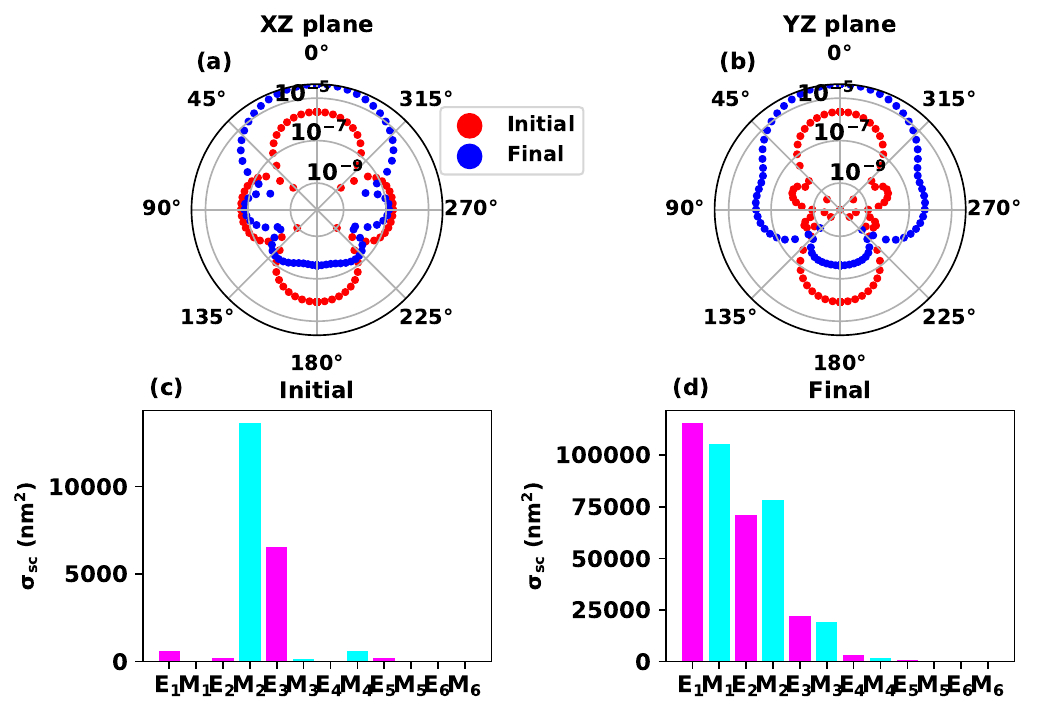} 
  \caption{Top: Polar plots of the radiation pattern in the $XZ$ (a) and $YZ$ planes (b). Bottom: Multipole contributions to the scattered cross-section of the initial (c) and final arrangement (d). For visual clarity, bars representing electric multipole contributions are shown in magenta, while magnetic multipoles are in cyan. All results are shown at the design wavelength.}
\label{sca2}\end{figure} 

The optimized structure exhibits a pattern obtained in alternative optimizations with other initial arrangements as well, with some spheres shifted along the $z$-direction. Generally, it is observed that the spheres move closer to each other and increase in size. The initial arrangement is 6-fold symmetric. However, the illumination breaks the symmetry for the gradients except for the mirror symmetry. For this reason, the final arrangement also possesses this symmetry. However, since the target scattering pattern is not symmetric about the $z$-axis, the optimized structure necessarily breaks the initial $6-$fold rotational symmetry by introducing unequal displacements along $z$. The total scattering is also increased, as expected given the larger sphere sizes in the final arrangement. 

To illustrate the changes, the radiation pattern and multipole decomposition for the initial and final arrangements at the design wavelength are shown in Fig.~\ref{sca2}. The radiation pattern, plotted here in log scale in Fig.~\ref{sca2}~(a) and (b), exhibits differences between the two considered planes, particularly in the strength of the sidelobes. Both radiation patterns indicate that the higher final ratio was achieved by a reduction in backward scattering and a simultaneous increase in forward scattering. 

The multipole decomposition of the scattering response at the design wavelength as shown in Fig.~\ref{sca2}~(c) and (d) was performed by expanding the final T-matrix about a single origin, necessitating the inclusion of higher-order multipoles, i.e., this is the response expressed in the global T-matrix. In the initial arrangement, total scattering is low and dominated by two multipolar components: magnetic quadrupole (61\%) and electric octupole (30\%) (Fig.~\ref{sca2}~(c)). The increased scattering strength in the optimized structure is achieved by enhancing the contribution of individual multipoles—gradually, decreasing in strength for higher orders, and effectively balancing the electric and magnetic components of the same order (dipoles: 28\% and 25.2\%, quadrupoles: 17\% and 18.5\%, octupoles: 5.3\%  and 4.5\%), as shown in Fig.~\ref{sca2}~(d). The pattern is consistent with the common understanding of the Kerker effect. Specifically, the different electric and magnetic multipoles contribute comparably to the total scattering response. Thanks to the suitable interference, the scattered light directly in the forward direction is two orders of magnitude larger than the light directly scattered in the backward direction. We note that, in general, equal contributions from electric and magnetic multipoles are not required~\cite {alaee2015generalized, liu2018generalized}.

Finally, we verify the functionality of the designed structures in dedicated experiments. To perform the experiments, we chose microwave frequencies. This range is experimentally less demanding, while the same arrangement can, in principle, be implemented at optical frequencies, provided suitable materials and nanofabrication methods are used. To prepare the experiments, the optimization was repeated using the experimentally measured material parameters of ABS plastic, which was chosen as the material from which the spheres were made. The target wavelength was \SI{10}{\centi\meter}, corresponding to a target frequency of 3~GHz. The same initial cluster arrangement was used as the starting point for the optimization, and the result of the optimization was very similar to the design introduced above. Both the initial and optimized arrangements were fabricated using 3D-printing and encapsulated in foam. The foam has electromagnetic properties close to air at the operating band and serves as a mechanical scaffold that maintains the spheres at their designed positions. With that, we can assure that the spheres are arranged as designed and structurally stable.

\begin{figure}[htbp]
  \centering
  \begin{subfigure}[b]{0.48\linewidth}
    \centering
    \includegraphics[width=\linewidth]{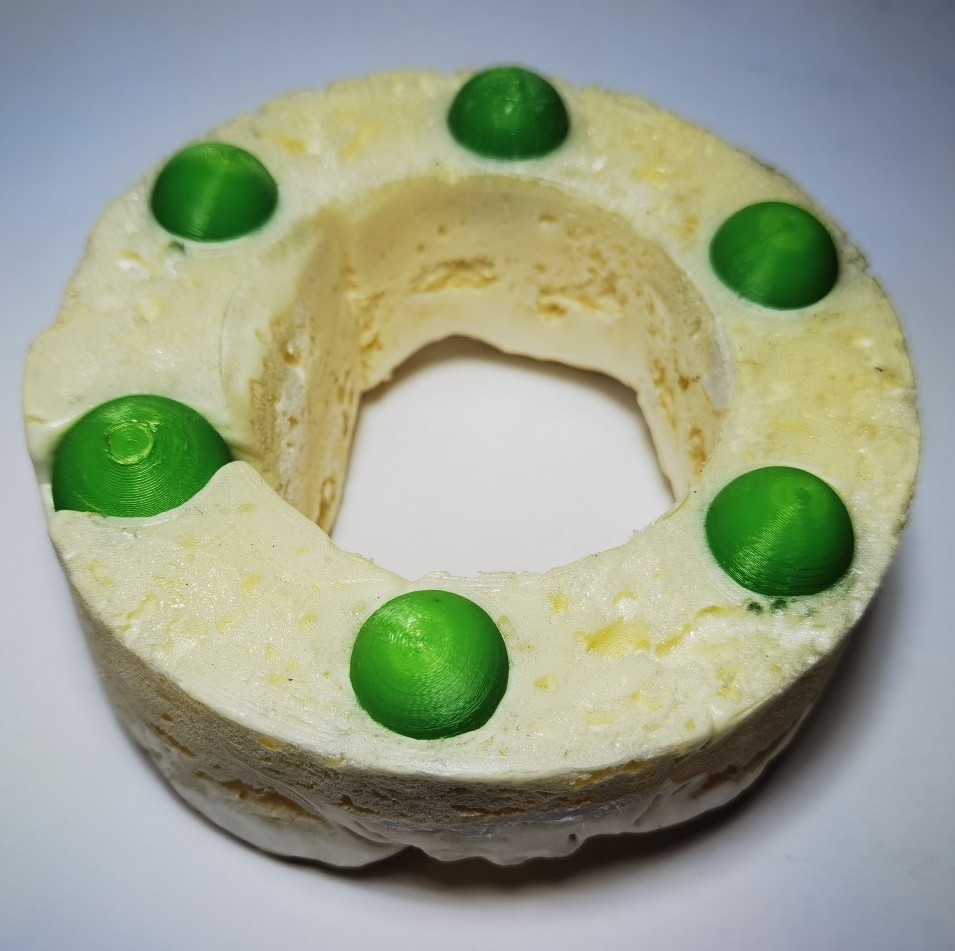}
    \caption{}\label{expa}
  \end{subfigure}\hfill
  \begin{subfigure}[b]{0.48\linewidth}
    \centering
    \includegraphics[width=\linewidth]{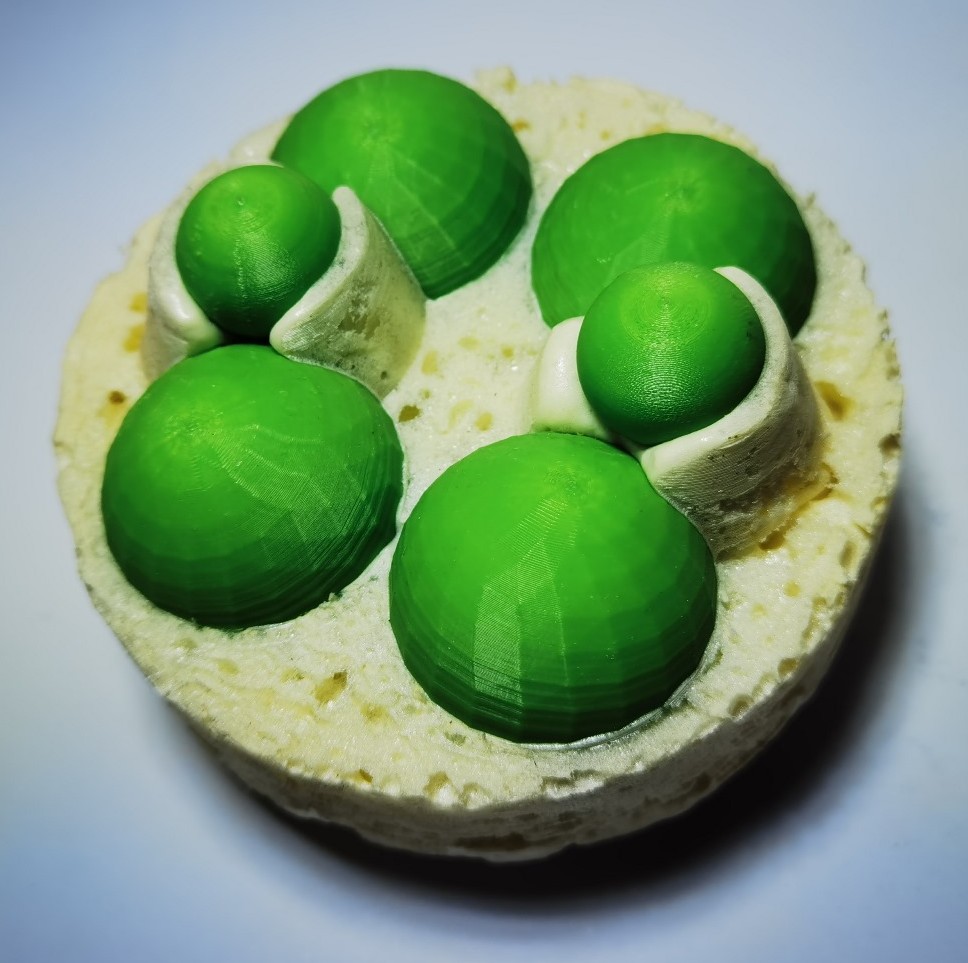} 
    \caption{}\label{expb}
  \end{subfigure}

  \vspace{0.5cm}

  \begin{subfigure}[b]{0.48\linewidth}
    \centering
    \includegraphics[width=\linewidth]{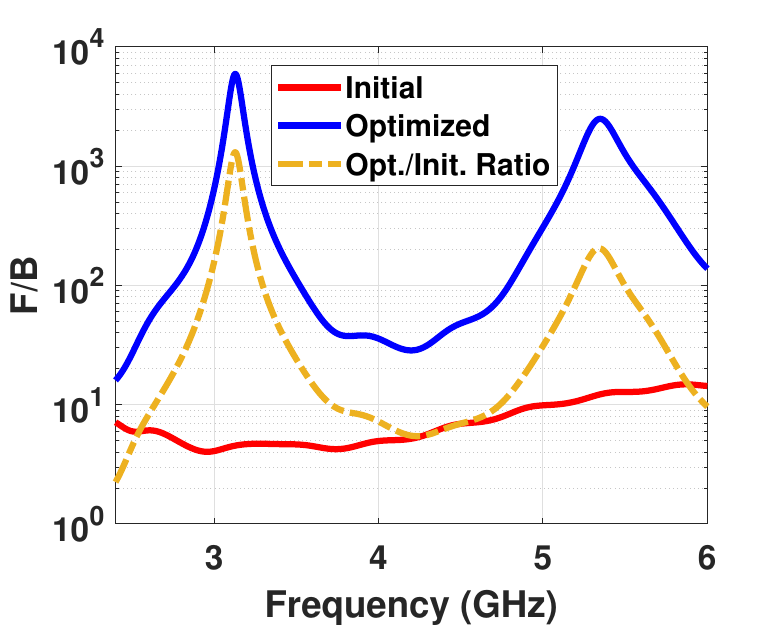}
    \caption{}\label{expc}
  \end{subfigure}\hfill
  \begin{subfigure}[b]{0.48\linewidth}
    \centering
    \includegraphics[width=\linewidth]{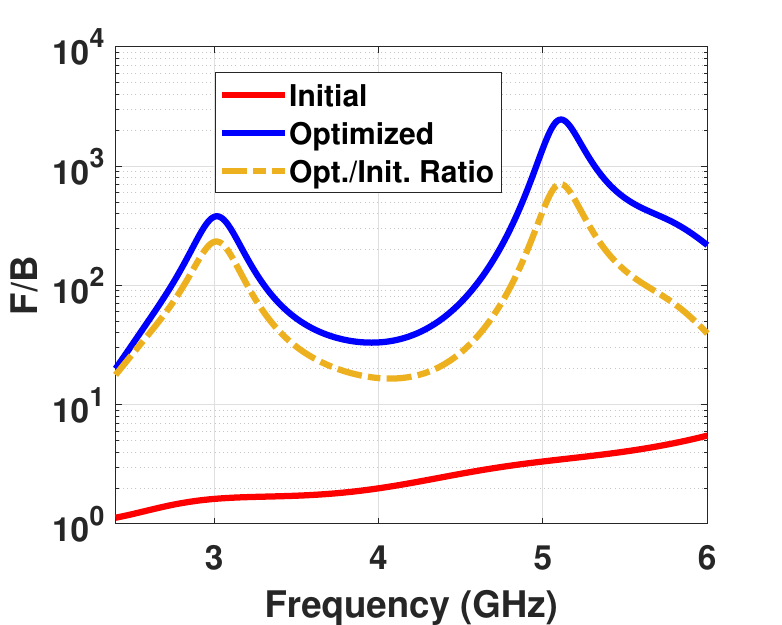}
    \caption{}\label{expd}
  \end{subfigure}

  \caption{Fabricated samples using 3D-printing and foam
hosting: (a) the initial cluster and (b) optimized cluster. The ratio of directly forward (0$^\circ$) to directly backward (180$^\circ$) scattering under $y$-polarized plane wave for both structures -- initial (solid blue) and optimized
(solid red), and their quotient (optimized/initial, dash-dotted): (c) experimental measurements and (d) CST numerical simulations.} \label{experiment}
\end{figure}

The electromagnetic scattering responses of the clusters were measured in the direct forward and backward directions, with the results shown in~Fig.~\ref{experiment}\subref{expc}. The simulation results with CST Studio Suite are demonstrated as a cross-check in Fig.~\ref{experiment}\subref{expd}. Generally, good agreement is observed, with a tremendously enhanced forward-to-backward scattering ratio of $10^2$ or more at or near the design frequency. Slight deviations occur, nevertheless, between experiment and simulations. After testing the sensitivity of the response to uncertainties in some selected parameters, we attribute the small frequency shift and larger value of the observed scattering ratio peak to minor uncertainties in the positions of the spheres, rather than to inconsistencies in the material properties.
Further details of the experimental realization, along with more measurement results and a sensitivity analysis, are provided in the Supplementary Material. Nevertheless, we can conclude from these experiments that the inverse designs we provide can be translated into tangible technology.

\subsection{Metasurface with complex unit cell} 
Next, we discuss as a second example the inverse design of a metasurface that consists of a periodic arrangement with a complex unit cell. Each unit cell contains five spheres. Our purpose is to achieve a strongly polarization-dependent reflection. As discussed previously~\cite{babicheva2017resonant}, the optical response of individual scatterers is modified by the lattice interaction, and the overlap of multipole contributions with different parity can cancel light propagation in one direction. Now, if the unit cell breaks 4-fold symmetry, it is possible to achieve this cancellation effect for only one of the polarizations of the incident plane wave. 

The objective function is defined as:
\begin{equation}
   \mathrm{FOM}  = \left| R_{y} - R_{x} \right|\,,
\end{equation}
where $R$ denotes the total reflectance, and the subscript indicates the $x$- or $y$-polarization of the incident light.
\begin{figure}[ht] 
  \centering
  \includegraphics[width=\linewidth]{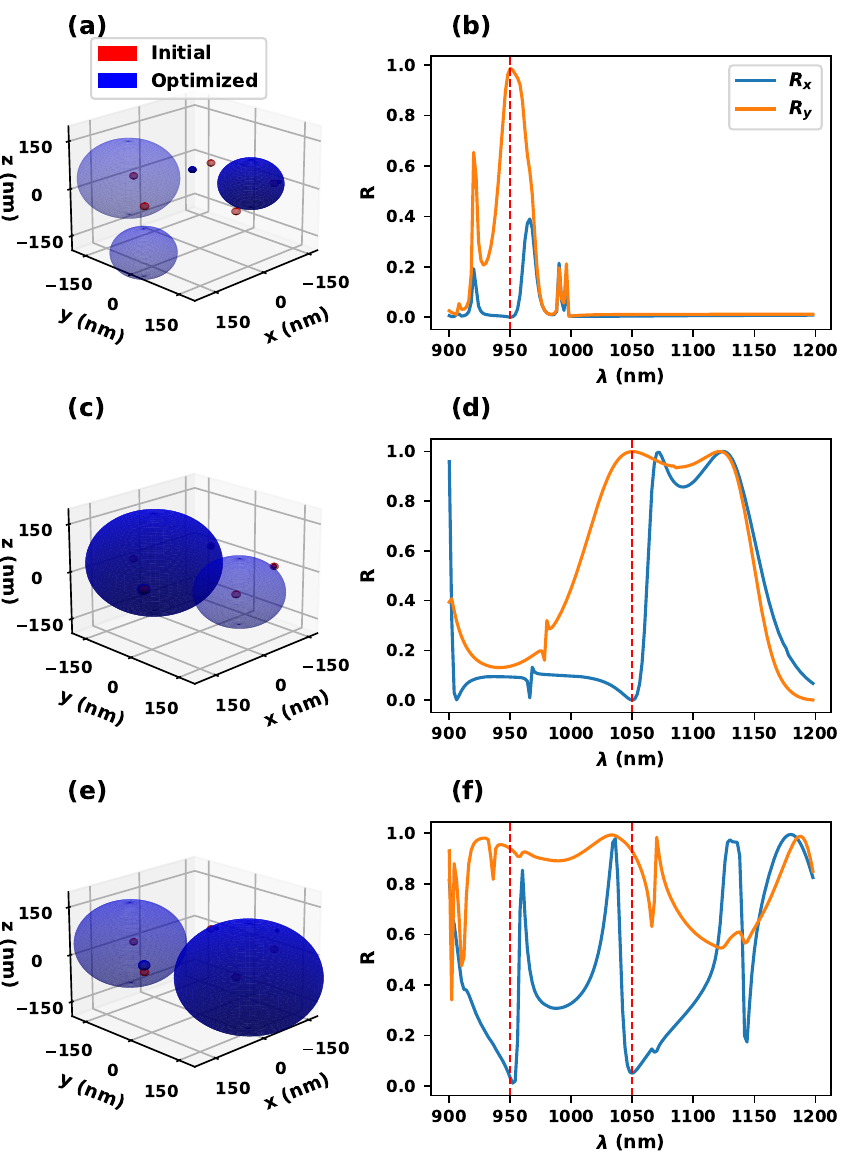}
  \caption{Initial (red) and optimized (blue) constituents of the metasurface unit cell targeting reflection peak for one linear polarization of the incident light and a low reflection for the other one at (a) 950~nm, (b) 1050~nm, and (c) simultaneously 950~nm and 1050~nm. The transparency of each sphere decreases with its proximity to the viewer. Panels (d--f) show corresponding reflection spectra for $x$-polarized (blue curves) and $y$-polarized (orange curves) incident light. Vertical red dashed lines indicate the target wavelength(s).}
\label{Metasurface}\end{figure}

Clearly, when setting the initial arrangement, the symmetries can be broken or preserved. We choose an odd number of spheres, five in total, and position them in a circle such that one sphere lies on the $x$-axis. This way, the unit cell is symmetric with respect to the $x$-axis, but not to the $y$-axis. Hence, we can expect the cancellation to occur for the $x$-polarized incident field. 

The setup parameters include the initial radius of a single sphere and the radius of the circle along which the spheres are arranged. The example considers silicon spheres with relative permittivity values taken from~\cite{schinke2015uncertainty} embedded in a medium with a relative permittivity of 2.25. Each sphere of the initial arrangement has a radius of 10~nm, the radius of the circle is set to 170~nm, and the pitch is 600~nm. The small size of the initial spheres is chosen to ensure ample free space for the scatterers to move and grow during the optimization process. The design wavelengths selected below exceed the unit cell size, ensuring operation within the non-diffractive regime. The incident field is always a plane wave propagating along the $z$-axis (normal incidence), and it is either $x$- or $y$-polarized.

To tune the resonance, we leave the radii and positions of the independent scatterers in the unit cell as free parameters while keeping the lattice constant fixed in the optimization. This corresponds to $4N=20$ parameters The spheres are not allowed to overlap and shrink below 5~nm, as in the previous example. The no-overlap condition was modified to enforce an additional minimal separation of 5~nm between spheres, based on the observation that they tend to approach each other at subnanometer distances. An additional constraint restricts the position of the spheres within the circumscribing sphere of the unit cell, since the T-matrix of the arrangement in the unit cell is expanded about a common origin. Multipole contributions up to 7th order are included during optimization, and the final arrangements are recalculated up to 15th order. The splitting parameter $\eta$ in the Ewald summation is selected automatically using a \texttt{treams} routine based on the wavenumber and lattice constant, and for our setup, the values are 0.42 and 0.46 for 950 nm and 1050 nm wavelengths, respectively. All optimizations run for up to 250 iterations and reach convergence within this number of steps. 
 
From optimization at different wavelengths, we note that the optimizer can reach the high reflectance value for many wavelengths within a particular wavelength range, and we demonstrate the optimization results obtained at 950 and 1050~nm in Fig.~\ref{Metasurface}(a-b) and (d-f). 

To explore a more challenging design, we also focus on achieving dual-band performance. It is realized by applying an adaptive weighting scheme to balance the optical response at two wavelengths. The objective is defined as follows: 

\begin{align}
f_i &= \bigl\lvert R_y(\lambda_i) - R_x(\lambda_i)\bigr\rvert\, ,\quad i=1,2\\
\mathrm{FOM} &= \frac{2f_1 f_2 + \varepsilon \min(f_1, f_2)}{f_1 + f_2 + \varepsilon}\, , 
\end{align}
where $\varepsilon$ is a small regularization parameter. The regularization ensures that the term with the smaller value is given greater emphasis.  
It can be observed in Fig.~\ref{Metasurface}~(f) that for the incident plane wave with $x$-polarization sharp peaks are produced, while a plane wave with $y$-polarization is responsible for the wide resonance curve in reflectance. 

\begin{figure}[ht] 
  \centering
    \includegraphics[width=\linewidth]{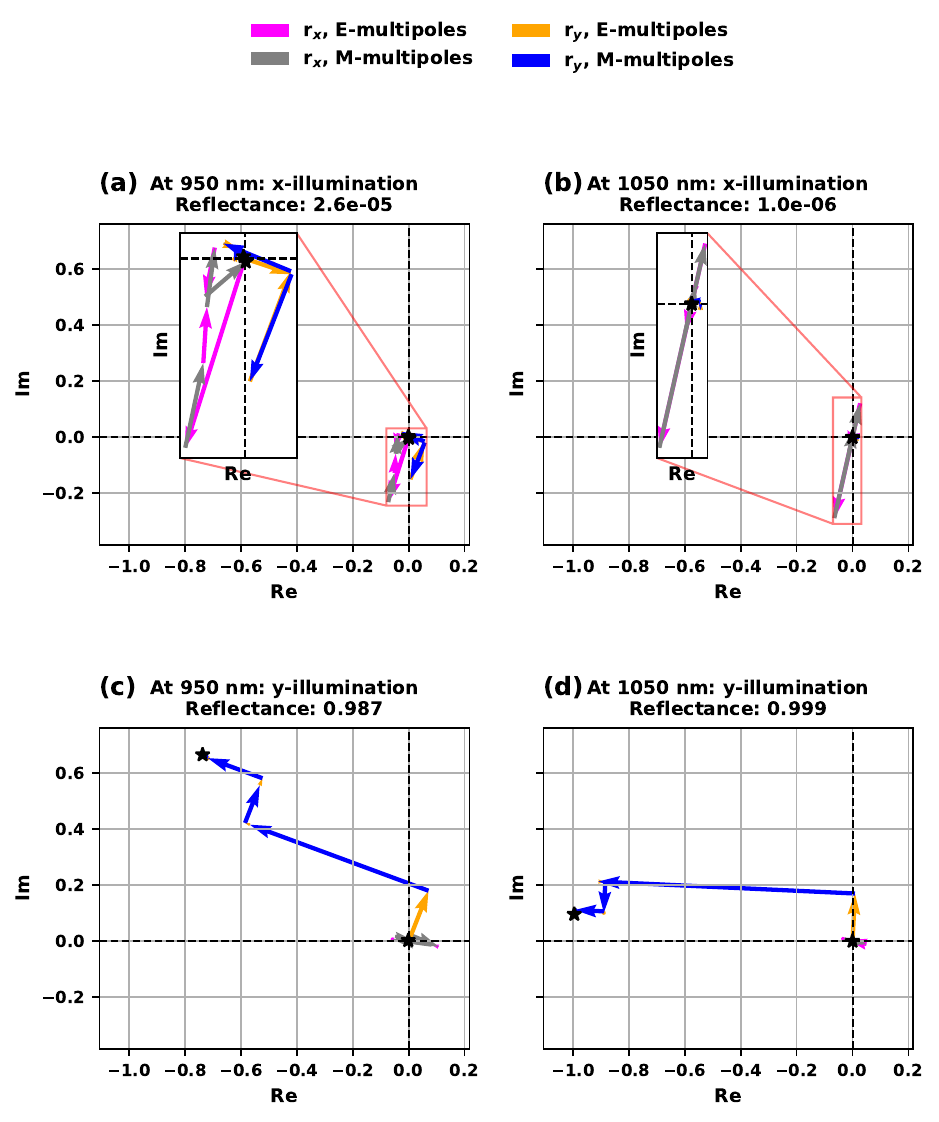}
  \caption{Phasor diagram of contributions of electric and magnetic multipoles to the backward-scattered field coefficients.  Panels (a, c) show the results for a non-diffractive metasurface optimized at 950~nm, while panels (b, d) demonstrate the results of optimization at 1050~nm. Starting from 0, the vectors build up as $E_1$,  $M_1$, $E_2$,  $M_2$, and so on, where $E$ stands for electric and $M$ for magnetic multipoles, followed by the multipole order. Each panel shows the phasor build-up of $x$- and $y$-polarized scattered field components separately. However, the cross-polarized components are negligible.  For both optimized structures illuminated with $x$-polarized light, in a) and b), the electric and magnetic multipole contributions cancel each other, whereas in c) and d), under $y$-illumination, they sum up to a high value of reflectance.}
\label{lat_single} \end{figure}

\begin{figure}[tbp] 
  \centering
  \includegraphics[width=\linewidth]{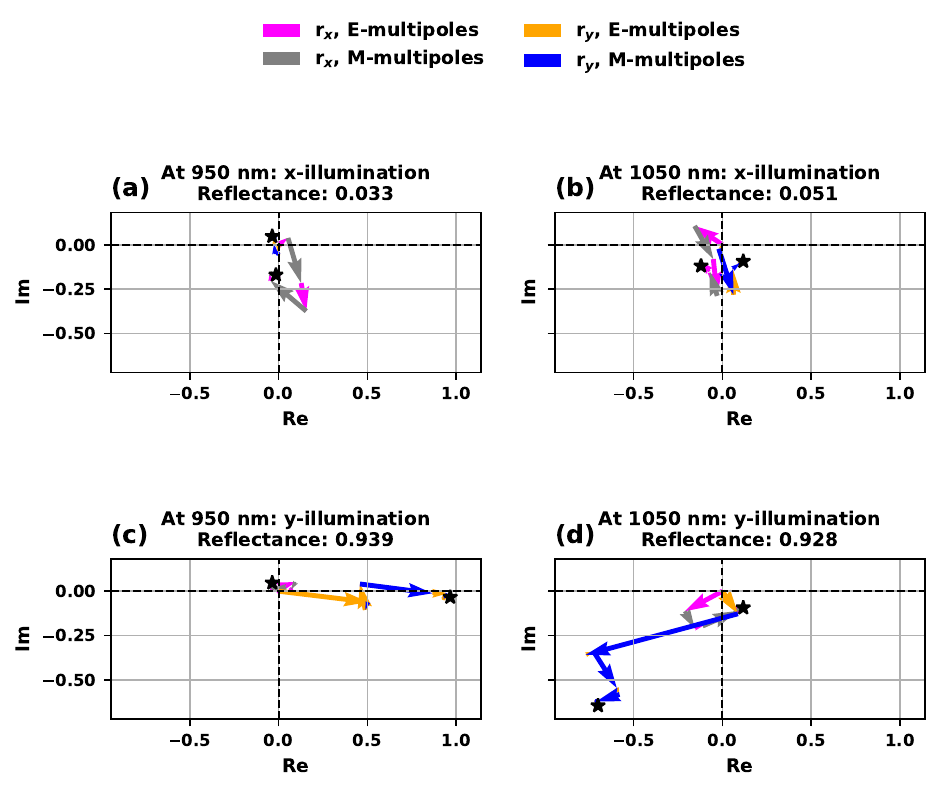}
  \caption{Phasor diagram of contributions of electric and magnetic multipoles to the complex reflection coefficient. The non-diffractive metasurface was optimized for maximum reflectance contrast at simultaneously 950~nm and 1050~nm. Panels a) and b) demonstrate the contributions of multipoles to the complex reflection coefficient for $x$- and $y$-components of the scattered field, when the metasurface is illuminated by $x$-polarized plane wave, which results in minimal final reflectance. In panels c) and d), the incident plane wave is $y$-polarized, and the final reflectance is high. Each panel shows the phasor build-up of $x$- and $y$-polarized scattered field components separately. }
\label{lat_multi} \end{figure}

For all optimizations, we observe that the number of spheres is effectively reduced, as the significant response is obtained only from two to three spheres that have increased considerably in size. This is to be expected given the limited space available during optimization. Optimizations that omit the shrunken spheres from the initial configuration reproduce the same final arrangement. 

To investigate the underlying behavior of the T-matrix, the phasor diagrams of the contributions of different multipoles to the components of the complex reflection coefficient are shown in Fig.~\ref{lat_single} for the single wavelength optimized geometries, and in Fig.~\ref{lat_multi} for the dual-wavelength optimized geometry. We work with T-matrices that already include lattice interactions. As the coefficients that contribute to the backward-propagating field are needed, we switch to plane-wave basis quantities for clarity.

In Appendix~\ref{appendix:translation}, exact formulas are provided to compute the expansion coefficient in a plane wave basis knowing the expansion coefficients in the spherical wave basis, which involve a summation procedure over all orders $l$ and $m$.
Rather than performing the entire summation at once, we separate the sum over particular multipole orders and polarizations, and derive the corresponding partial S-matrices, each accounting only for electric dipole, magnetic dipole, and so on.  Next, the complex-valued reflection coefficients are calculated by multiplying the $S_{\downarrow \uparrow}$ block with the vector representing the expansion coefficients of the plane wave, which contains one and zero according to the polarization.
The final reflectance is: 
\begin{equation} \mathrm{R} = \left| \mathrm  \sum_{i=1}^N{ \mathrm r_i} \right|  ^2\, ,
\end{equation}
where $N = 2 l_{\mathrm{max}}$. The reflection coefficients corresponding to each multipole with a definite total angular momentum and parity are depicted as phasors, and their vector sum is the total reflection coefficient. Although the expansion includes 15 multipole orders, the phasor diagram shows that only up to octupole order contributions are significant. 

In Fig.~\ref{lat_single}, the electric and magnetic multipoles cancel each other almost perfectly under $x$-polarized light. For $y$-illumination, the contributions add up to nearly perfect reflectance. The cross-polarized terms are negligible. Meanwhile, for dual-wavelength optimization in Fig.~\ref{lat_multi}, the situation is not the same. 
We observe a few strong multipole contributions under $y$-polarized illumination that are not compensated, as expected. For the $x$-polarized incident field, there are components of smaller magnitude, with some of the phasors directed in opposite directions. However, there is no explicit pairwise cancellation characteristic of the Kerker effect. In both illumination scenarios, the cross-polarized field components are nonzero and of similar magnitude, indicating that this can be disregarded when seeking the actual reason for the difference in outcomes. Clearly, the dual-band objective is more complicated to satisfy, and one cannot achieve total cancellation; however, the optimization procedure found a sufficiently close solution, and the reflectance for $x$-polarized light can be considered negligible.

The demonstrated results reinforce the conclusion that gradient-based methods are a powerful tool for optimization. As for all local optimization techniques, the achievable results can be sensitive to the choice of initial conditions. Common approaches to avoid convergence to suboptimal local maxima include combining these techniques with global optimization methods or performing multiple runs with varied initial conditions. Nevertheless, the gradient-based framework is essential for fine-tuning design parameters.

\section{Conclusion}

We have developed and evaluated a differentiable computational framework for the inverse design of nanophotonic systems using the T-matrix method. The Python code is available under the MIT license (\url{https://github.com/tfp-photonics/dreams}).
We applied this framework to design various structures consisting of elaborate arrangements of spherical scatterers in finite clusters and on periodic lattices. A key strength of our framework is its ability to simultaneously optimize the positions and radii of spheres, leveraging strong multipolar interactions. This capability extends directly to the positions of arbitrarily shaped scatterers, and the method can be used together with other differentiable solvers to compute derivatives with respect to any geometrical parameter. The required derivatives are efficiently obtained through automatic differentiation. Even a moderate number of parameters generates an ample design space, making brute-force parameter sweeps and some global optimization techniques infeasible. 

Specifically, we demonstrate a maximized forward-to-backward scattering ratio for a cluster of spheres and a tailored polarization-dependent reflectance for a metasurface with a complex unit cell.
Both case studies involve the cancellation of scattered fields in a specific direction, exemplifying the generalized Kerker effect. The comprehensive T-matrix approach easily reveals the contribution of different multipoles to the optimized scattered field. We ensure that the simulation method itself does not impose restrictions on the inclusion of higher-order multipolar components in the optical response. From the optimization curves, we observe that the optimizations start from comparatively low initial values and converge to high-performance solutions that respect all constraints. Combining multiple objectives, as seen in the example of dual-wavelength optimization, achieves the desired functionality, even if it does not strictly follow the expected underlying mechanism. This further enhances the flexibility of the design framework. 

This work opens new possibilities for the automated, high-fidelity design of complex nanophotonic architectures where multipolar interference plays a central role. By integrating differentiable programming with the T-matrix formalism, it lays the groundwork for scalable, mechanism-agnostic optimization across diverse application domains, from directional scattering to metasurface engineering. The framework is well-positioned to accelerate the inverse design of functional photonic systems beyond traditional heuristic-guided approaches.

Finally, we wish to add that at the time of the submission, we became aware of a manuscript developing similar ideas~\cite{jackson2026pymiediff}. This underpins the timeliness and importance of the differentiable formulation of algorithms that solve multiple-scattering problems, a key requirement for inverse design of photonic nanostructures.

\section*{Supplementary Material}
The supplementary material presents gradient validation and scaling analysis in terms of scatterer number and multipolar order. It also compares the approach with gradient-free optimization and gradient-based optimization iterations using full-wave solvers, and describes the microwave experiments, including sample fabrication, characterization techniques, and sensitivity analysis.

\begin{acknowledgments}
N.A. and C.R. acknowledge financial support by the Federal Ministry of Research, Technology and Space (BMFTR) within the project DAPHONA
(16DKWN039). J.D.F. and C.R. acknowledge financial support by the Helmholtz Association in the framework of the innovation platform “Solar TAP”. O.K. and C.R. acknowledge support from the German Research Foundation within the Excellence Cluster 3D Matter Made to Order (EXC 2082/2 under project number 390761711) and by the Carl Zeiss Foundation. N.A. acknowledges support from the Max Planck School of Photonics (MPSP). N.A. and J.D.F. acknowledge support from the Karlsruhe School of Optics and Photonics (KSOP).
The work of D.V. was supported by the 1.1.1.9 Activity "Post-doctoral Research" Research application No 1.1.1.9/LZP/1/24/166 "Linear Industrial Monitoring System based on Hyperspectral Cameras and AI Algorithms (LIF-HYCAI)". 
TAU Team acknowledges Israel Science Foundation (ISF Grant Number 1115/23). We acknowledge support by the KIT-Publication Fund of the Karlsruhe Institute of Technology.

\end{acknowledgments}

\section*{Data Availability Statement}

The data that support the findings of this study are openly available at~\cite{dreams}.

\appendix

\bibliography{aipsamp}

\section{Custom derivative}

\label{appendix:derivation}
The derivative of the spherical Bessel (and Hankel) functions is calculated using the recurrence relation. The special case of zero argument has to be treated accordingly. For $v>1$, it can be set to zero, while for $v=1$, the Taylor series expansion of the $sinc$ gives a non-zero number.  The code snippet is provided in Listing~\ref{lst:spherical_jn}.
\begin{lstlisting}[caption={Example: Custom spherical Bessel function in JAX}, label={lst:spherical_jn}]
def spherical_jn(v, z):
    z = z.astype(np.result_type(complex, z.dtype))

    return jax.pure_callback(
        lambda v, z: mod_jn(v, z).astype(z.dtype),
        jax.ShapeDtypeStruct(
            shape=np.broadcast_shapes(v.shape, z.shape),
            dtype=z.dtype
        ),
        v, z,
        vmap_method="legacy_vectorized"
    )

@spherical_jn.defjvp
def _jv_jvp(primals, tangents):
    """
    Custom JVP rule for spherical Bessel function.
    """
    v, z = primals
    _, z_dot = tangents  # v_dot = 0 since v is integer.

    jv_v_z = spherical_jn(v, z)
    jv_plus_1 = spherical_jn(v + 1, z)

    small_z = np.abs(z) < 1e-8

    djv_dz = np.where(
        (small_z & (v != 1)), 0.0j,
        np.where(
            (small_z & (v == 1)), 1/3,
            v * jv_v_z / z - jv_plus_1
        )
    )

    return jv_v_z, z_dot * djv_dz
\end{lstlisting}

\section{Translation from periodic vector spherical wave coefficients to plane wave coefficients}
\label{appendix:translation}
The scattered field summed over the periodic lattice in the plane wave basis can be written as follows:
\begin{equation}
\begin{aligned}
\mathbf{E}_{\rm scat}^{(\mathrm{lattice}),\,d}(\mathbf r)
&= 
\sum_{g\in\Lambda^*}
e^{\,i(\mathbf k_\parallel + g)\cdot \mathbf r}
\Bigl[
  C_{\rm TE}(g)\,\hat e_{\rm TE}(\mathbf k_\parallel + g,\,d)\\
&\quad\;\;+\;C_{\rm TM}(g)\,\hat e_{\rm TM}(\mathbf k_\parallel + g,\,d)
\Bigr].
\end{aligned} 
\end{equation}
In the expression above, $\mathbf k_\parallel$ is the tangential component of the plane wave, and $g$ is the diffraction order. The expansion coefficients are given by:
\begin{equation}
\begin{pmatrix}C_{\rm TE}(g)\\[4pt]C_{\rm TM}(g)\end{pmatrix}
=
\sum_{\ell=1}^{\ell_{\max}}
\sum_{m=-\ell}^{\ell}
F_{\ell m}(g)\,
\underbrace{\begin{pmatrix}
  \tau_{\ell m}(\theta_g) & \pi_{\ell m}(\theta_g)\\[4pt]
  \pi_{\ell m}(\theta_g)  & \tau_{\ell m}(\theta_g)
\end{pmatrix}}_{A_{\ell m}(\theta_g)}
\begin{pmatrix}p_{e,\ell m}\\[4pt]p_{m,\ell m}\end{pmatrix},
\end{equation}
where  $\pi_{\ell m}(\theta_g)$ and  $\tau_{\ell m}(\theta_g)$ are the two angle-dependent functions defined in terms of the associated Legendre functions $P_l^m$:
\begin{equation}
\pi_{l m}(\theta)
=\frac{m}{\sin\theta}\,P_l^m\bigl(\cos\theta\bigr),
\qquad
\tau_{l m}(\theta)
=\frac{d}{d\theta}\,P_l^m\bigl(\cos\theta\bigr).
\end{equation}
Finally,
\begin{equation}
\begin{aligned}
    F_{\ell m}(g)
\;&=\;
-\;\frac{i\pi\,N_{\ell m}}{\,a\,k\,i^{\,\ell-m}\,}\;
\frac{1}{\sqrt{1 - \lvert \mathbf k_\parallel + g\rvert^2/k^2}}\,, 
\\
\cos\theta_g &= \frac{\sqrt{k^2 - \lvert\mathbf k_\parallel+g\rvert^2}}{k}.
\end{aligned}
\end{equation}

where $N_{\ell m}$ is the normalization constant of the VSW, and $a$ is the area of the unit cell. 
%

\clearpage

\renewcommand\thesection{S\arabic{section}}
\renewcommand\thesubsection{S\arabic{section}.\arabic{subsection}}
\renewcommand\thesubsubsection{S\arabic{section}.\arabic{subsection}.\arabic{subsubsection}}

\renewcommand\thefigure{S\arabic{figure}}
\renewcommand\thetable{S\arabic{table}}
\renewcommand\theequation{S\arabic{equation}}

\setcounter{section}{1}
\setcounter{subsection}{0}
\setcounter{subsubsection}{1}
\setcounter{figure}{1}
\setcounter{table}{1}
\setcounter{equation}{1}

\section*{ Supplementary Material}
The following aspects are discussed in dedicated sections:

\begin{enumerate}
    \item \hyperref[sec:Gradient validation]{\bf{Gradient validation}:} In this section, we validate the gradients that have been obtained from the automatic differentiation by comparing them to gradients obtained with a central-difference approximation.  
    \item \hyperref[sec:Scaling]{\bf{Scaling}:} In this section,  we quantify the computational costs of the inverse design and discuss the scaling behavior of our implementation both in terms of the number of scatterers and multipolar orders for a comprehensive assessment of the computational expenses.  
    \item \hyperref[sec:supp-other-methods]{\bf{Optimization using alternative methods}:} In this section, we compare the suitability of our approach for the inverse design to other existing techniques. Specifically, we distinguish between gradient-free optimization and gradient-based optimization using alternative full-wave solvers. 
    \item \hyperref[sec:Experimental realization at microwave frequencies]{\bf{Experimental realization at microwave frequencies}:} In this section, we present details of the experiments done in order to validate the suitability of the obtained designs. The experiments are conducted at microwave frequencies using dedicated samples, whose fabrication process we also described in depth, along with additional information on the characterization techniques used.
\end{enumerate}

\section{Gradient validation}
\label{sec:Gradient validation}
\begin{figure}[!hbt]
  \centering
  \includegraphics[width=\linewidth]{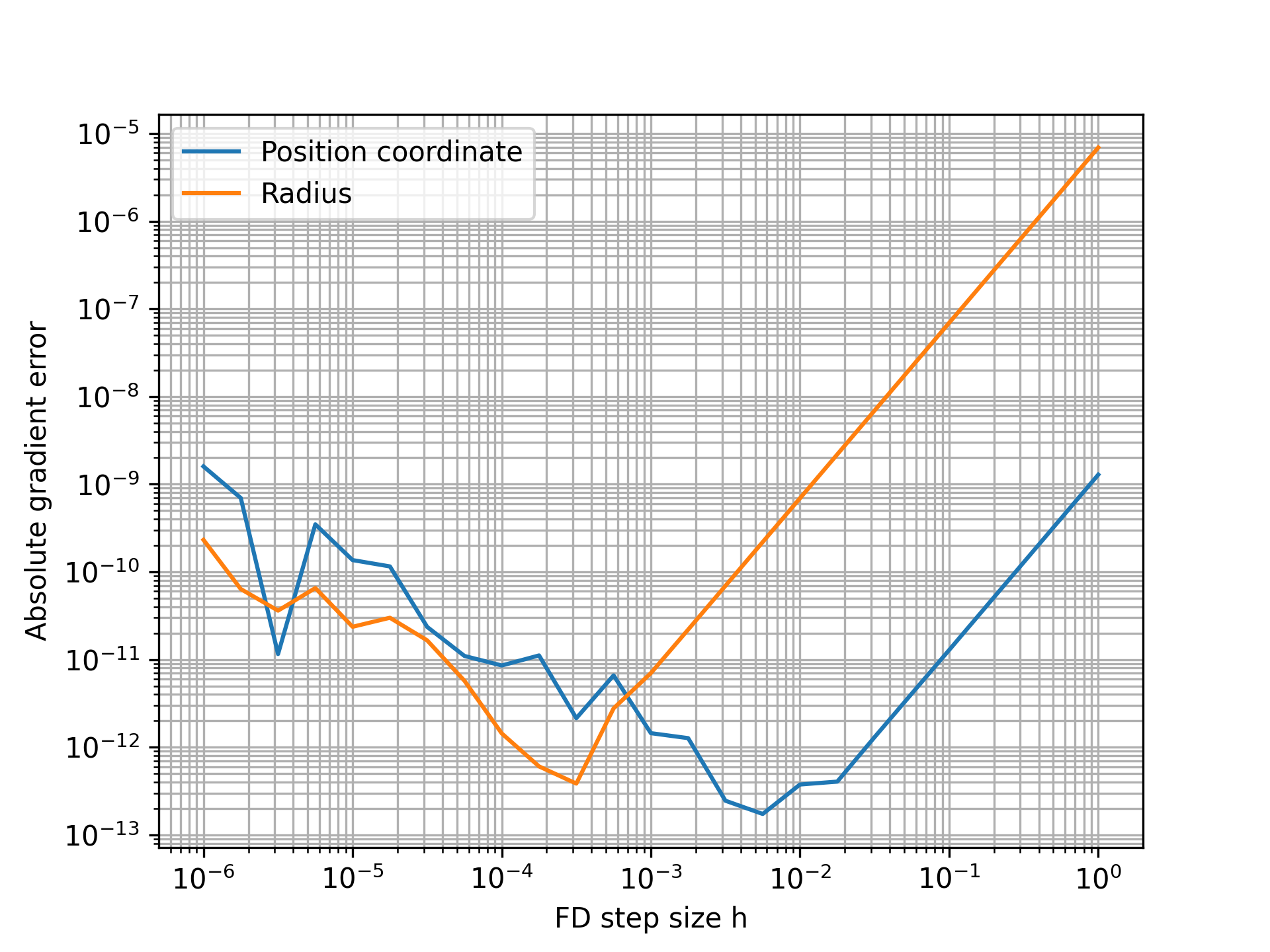}
  \caption{The absolute error between gradients obtained by automatic differentiation (AD) and by central finite differences (FD) for one position coordinate and one radius parameter as a function of the FD step size \(h\) (log--log scale).}
\label{grad}\end{figure}

To quantitatively validate the gradients computed with automatic differentiation, we compare them against gradients computed with finite differences for two parameters: one radius and one position coordinate. Here, we show the error between the gradients obtained with automatic differentiation and with central finite differences, i.e., for the objective of the initial arrangement discussed in the first example of the main text, $f(\mathbf{p})$, we compute via the central-difference approximation:
\begin{equation}
\left.\frac{\partial f}{\partial p_i}\right|_{\mathrm{FD}} \approx
\frac{f(\mathbf{p}+h\,\mathbf{e}_i)-f(\mathbf{p}-h\,\mathbf{e}_i)}{2h},
\end{equation}
and depict the absolute error
\begin{equation}
\left.\frac{\partial f}{\partial p_i}\right|_{\mathrm{FD}}-
\left.\frac{\partial f}{\partial p_i}\right|_{\mathrm{AD}}
\end{equation}
over a range of steps \(h\) in Fig.~\ref{grad}. As expected, in the finite-difference approach, large step sizes lead to truncation error (since we neglected higher-order Taylor terms), whereas for very small step sizes, the errors are dominated by numerical noise and round-off due to subtractive cancellation. Such an observation applies to both differentiation, i.e., with respect to the position coordinate and to the radius. The observations are a consistent example of the challenges that typically arise with finite difference schemes. 

\section{Scaling}
\label{sec:Scaling}

\textbf{Per-evaluation cost}
In this section, we quantify the computational costs of the proposed framework and its scaling with the problem size.

\begin{table}[!htbp]
\caption{Runtime per optimization iteration (wall-clock) for the two examples, measured after a one-time JIT compilation and averaged over 10 runs. $t_{\mathrm{fwd}}$ is one forward objective evaluation and $t_{\mathrm{fwd+bwd}}$ is a combined forward+backward pass including reverse-mode AD gradient computation. Peak memory is peak resident set size (RSS), measured as the maximum physical memory used by the process during the run.}
\label{tab:timing_examples}
\begin{ruledtabular}
\begin{tabular}{lccc}
Example & $t_{\mathrm{fwd}}$ (s) & $t_{\mathrm{fwd+bwd}}$ (s) & RSS (GB) \\
\hline
\begin{tabular}[c]{@{}l@{}}Ex.~1: finite cluster\\[-1pt]($N=6$, $l_{\max}=3$, $D=24$)\end{tabular}
& 3.58 & 7.68 & 1.83 \\
\begin{tabular}[c]{@{}l@{}}Ex.~2: metasurface\\[-1pt]($N=5$, $l_{\max}=7$, $D=20$)\end{tabular}
& 6.26 & 13.71 & 2.02 \\
\end{tabular}
\end{ruledtabular}
\end{table}

\begin{figure}[!htb]
  \includegraphics[width=\linewidth]{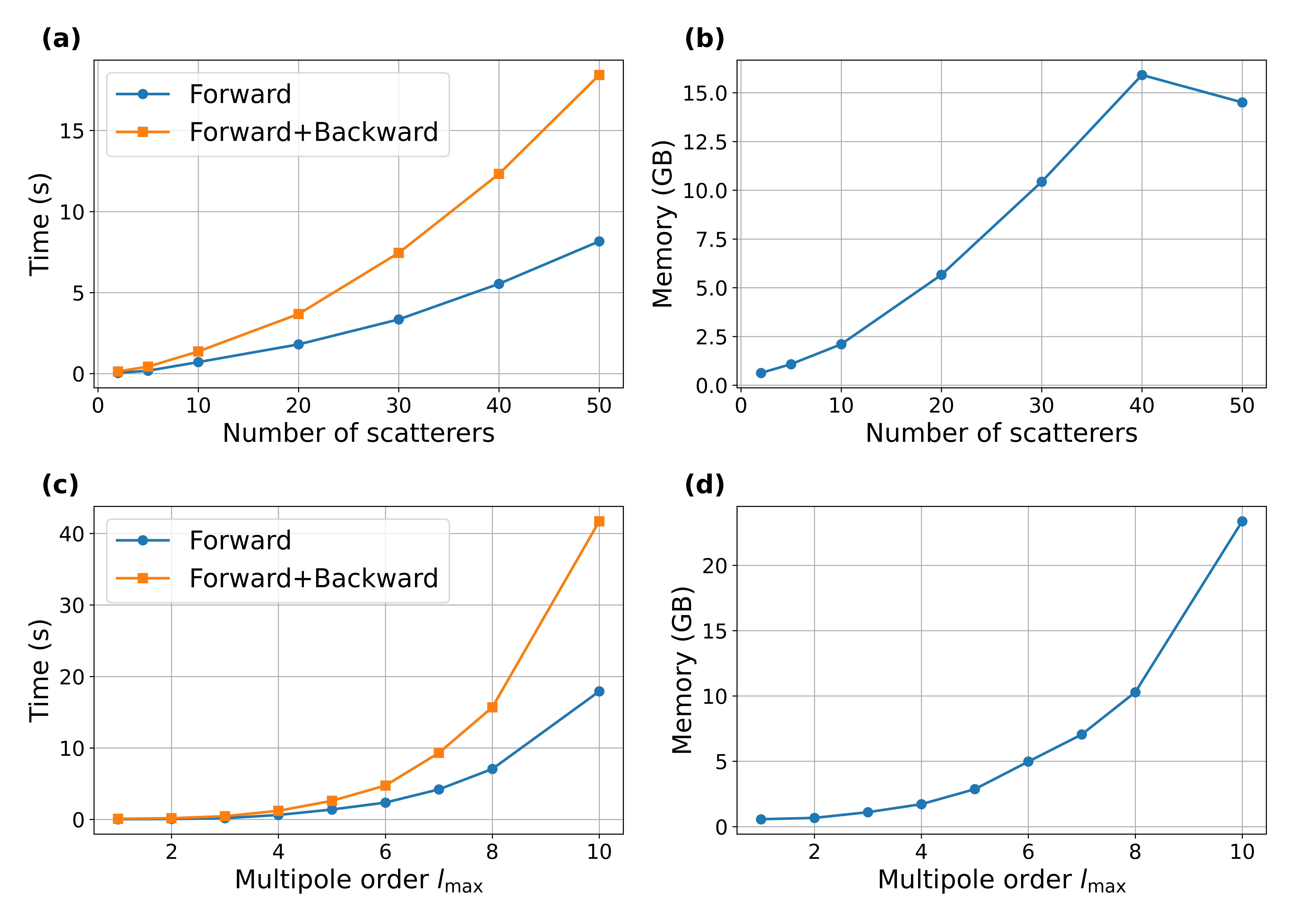}
  \caption{
  Wall-clock runtime and peak memory for a forward T-matrix evaluation and a combined forward and backward pass, where the backward pass corresponds to reverse-mode automatic differentiation. Memory values report the increase in process resident set size (RSS)  during execution on CPU.
(a) Runtime versus number of scatterers $N$ at fixed  $l_{\max}=3$. 
(b) Peak memory usage versus $N$ at fixed $l_{\max}=3$.
(c) Runtime versus multipole truncation order $l_{\max}$ at fixed $N=6$.
(d) Peak memory usage versus $l_{\max}$ at fixed $N=6$.
}
\label{scale}\end{figure}

\begin{figure}[!htb]
  \includegraphics[width=\linewidth]{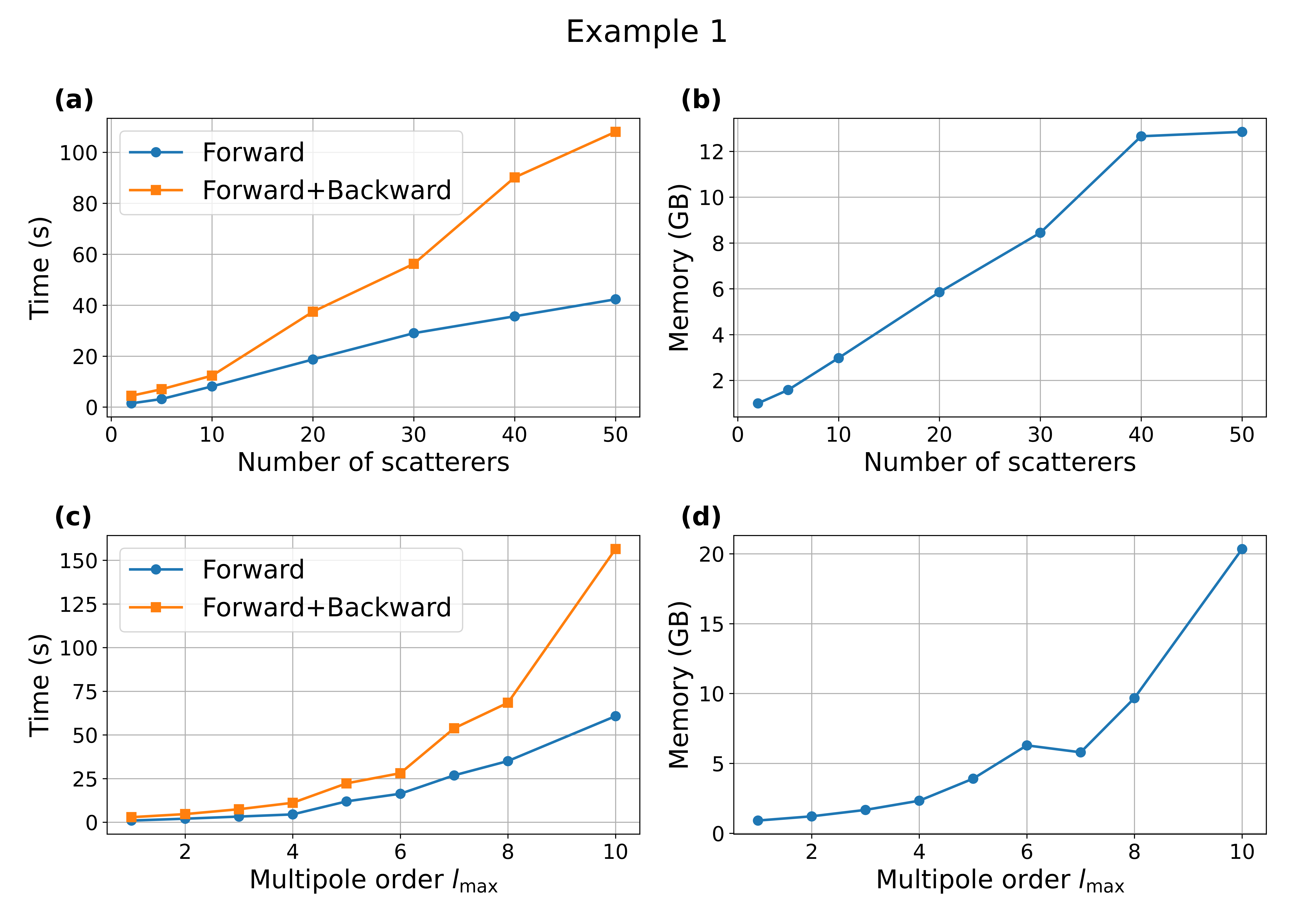}
  \caption{
    Wall-clock runtime and peak memory for the example of a finite cluster of spheres.  Memory values report the increase in process resident set size (RSS) during execution on CPU.  The objective is the ratio of the scattered power integrated over the forward and backward hemispheres.
(a) Wall-clock runtime versus number of scatterers $N$ for a forward evaluation and a combined forward and backward pass, where the backward pass corresponds to reverse-mode automatic differentiation. The objective is the ratio of the scattered power integrated over the forward and backward hemisphere.
(b) Peak memory usage versus $N$.
(c) Runtime versus multipole truncation order $l_{\max}$ at fixed $N$.
(d) Peak memory usage versus $l_{\max}$ at fixed $N$.
}
\label{scale_ex1}\end{figure}

\begin{figure}[!htb]
  \centering
  \includegraphics[width=\linewidth]{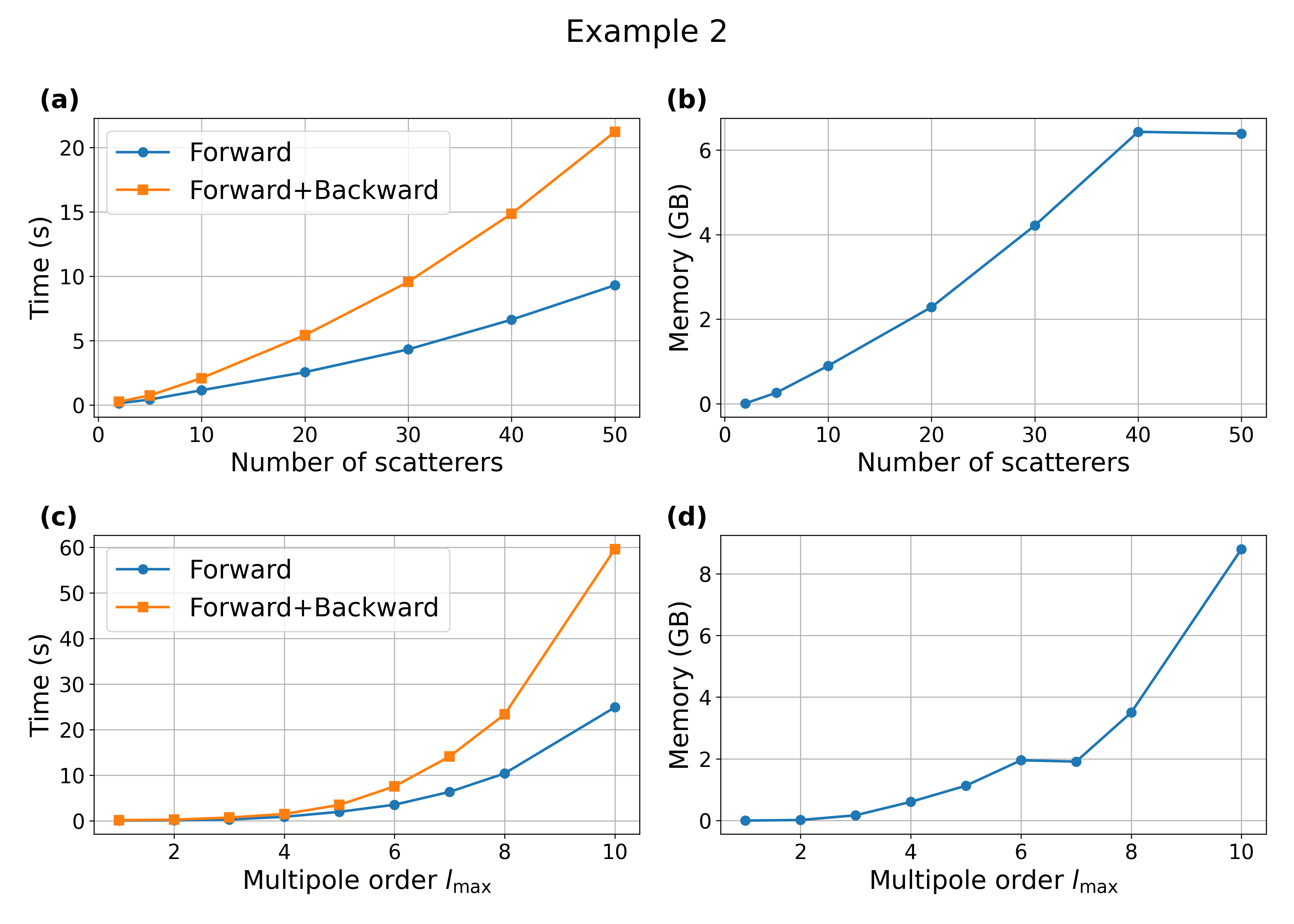}
  \caption{
  Wall-clock runtime and peak memory for the example of a metasurface with a complex unit cell. Memory values report the increase in process resident set size (RSS) during execution on the CPU. The objective is the difference between reflectances for the incident plane wave with orthogonal polarizations.
(a) Runtime versus number of scatterers $N$ at fixed $l_{\max}$.
(b) Peak memory usage versus $N$ at fixed $l_{\max}$. 
(c) Runtime versus multipole truncation order $l_{\max}$ at fixed $N$.
(d) Peak memory usage versus $l_{\max}$ at fixed $N$.
}
\label{scale_ex2}\end{figure}

For the two examples presented in the main text, we first report the computation time for a single optimization iteration in Table~\ref{tab:timing_examples}. Figure~\ref{scale} demonstrates the wall-clock runtime of the T-matrix computation and peak memory usage during execution, all measured after a one-time compilation, with varying the number of scatterers $N$ and the multipole truncation order $l_{\max}$. All measurements were performed on a dual-socket AMD EPYC 7413 system (48 cores / 96 threads), using \emph{float64} precision. Timings are averaged over 10 runs. The scaling behaviour for the actual objective of the first example is demonstrated in Fig.~\ref{scale_ex1}, and it is clear that explicit field evaluation over a very large grid of far-field points is quite an expensive computation. By contrast, for the per-evaluation calculation in the second example, we see in Fig.~\ref{scale_ex2} that the costs increase more moderately.

\begin{figure}[!htb]
  \centering
  \includegraphics[width=\linewidth]{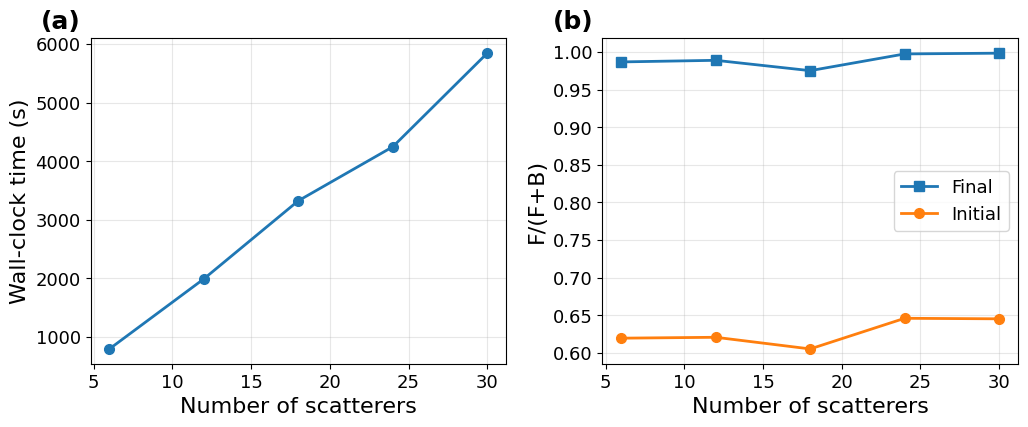}
  \caption{
End-to-end optimization performance as a function of the number of scatterers. (a) Total wall-clock time for the optimization, including one-time compilation overhead. In all cases, the optimization is run for 100 iterations. (b) Objective value at the initial configuration and after the optimization.
}
\label{scale-end}\end{figure}

\textbf{End-to-end optimization}
As an additional study, we perform a variation of the initial arrangement for the finite-cluster example. We take the initial ring of six spheres and add additional outer rings. For the forward-to-backward ratio, the increased overall size of the cluster also leads to an increased objective value. Therefore, we reformulate the objective as the fraction of the total scattered power directed into the forward hemisphere. The number of iterations is fixed to 100. We report the end-to-end optimization runtime, including the one-time compilation overhead, in Fig.~\ref{scale-end}(a). For the initial arrangement with all spheres positioned in the $z=0$ plane, the fraction of power scattered into the forward hemisphere is a bit above one-half. In all tested cases, closely reaching the maximum possible ratio is achievable, as can be seen in Fig.~\ref{scale-end}(b).

\section{Optimization using alternative methods}\label{sec:supp-other-methods}
\subsection{ Gradient-free optimization}
\begin{figure}[th]
  \centering
  \includegraphics[width=\linewidth]{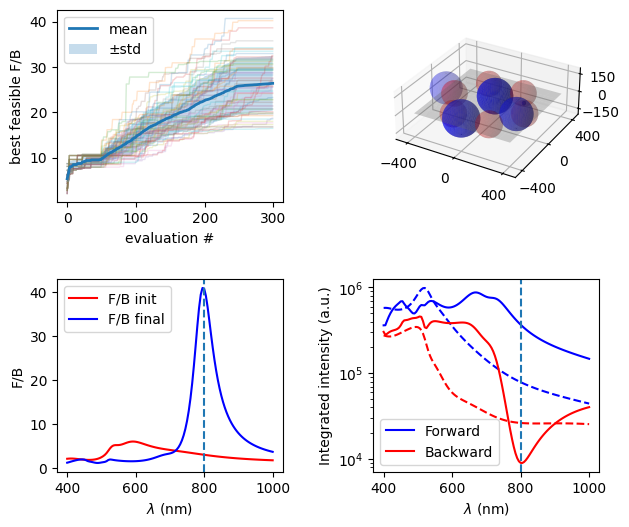}
\caption{Optimization of the forward-to-backward scattering ratio (F/B) for the finite cluster, best run among the 10 random initializations. \textbf{(a)}  Evolution of the objective value (F/B) over optimization steps for all optimization runs, comprising 10 random initializations for each considered hyperparameter setting. Faint curves show the individual optimization runs, while the solid line and shaded region indicate the mean and standard deviation across all runs. \textbf{(b)} Initial (red) and optimized (blue) sphere configurations. \textbf{(c)} F/B spectrum before (red) and after (blue) optimization, the vertical dashed line indicates the target wavelength. \textbf{(d)} The scattered
field intensity integrated over the forward (blue) and
backward (red) hemispheres. Dashed curves indicate the
initial arrangement and solid curves the optimized one. The
dashed vertical line marks the design wavelength.}
\label{bayes}\end{figure}

It is challenging to apply gradient-free methods to problems with tens of parameters while respecting all nonlinear constraints, which can lead to many rejected steps. For a comparison, we choose scalable constrained Bayesian optimization (SCBO), a constrained extension of Trust Region Bayesian Optimization (TuRBO), using the implementation provided in BoTorch~\cite{balandat2020botorch}. This approach was specifically directed at improving the efficiency of Bayesian optimization for high-dimensional optimization~\cite{eriksson2019scalable}, and had superior performance in a benchmarking paper~\cite{santoni2024comparison}. We set the same objective as in the first example of the main text with the finite cluster. The positions are parameterized by shifts relative to the positions on an initial ring with a radius of 300~nm. The maximum shifts in the xy-plane are 100~nm, and along the z-axis they are 120~nm. Radii of the spheres can range from 2~nm to 150~nm. For each arrangement, the constraint violation is also evaluated. The objective Gaussian Process (GP) surrogate was trained only on feasible samples, whereas the constraint GP was trained on all samples. We use ten random seed initializations with 50 initial feasible evaluations and 200 iterations. At each iteration, 4800 candidate points are generated within the trust region, from which the next evaluation is selected by constrained posterior sampling.
It is possible to vary the initial parameters to tune the behaviour of the optimization, such as the initial trust region size and the perturbation level (i.e., the number of coordinates that are changed per move). We varied the initial trust region size between 0.3, 0.5, and 0.8 in the normalized space for a fixed number of perturbation candidates, and the perturbation level between 5, 10, and 20 for a fixed initial trust region size. The best result that could be reached was 40.7, achieved with an initial trust region size 0.3 and 10 perturbed coordinates. The corresponding arrangements and optical characteristics are depicted in Fig.~\ref{bayes}. We note that, if the final arrangement obtained by gradient-based optimization were used as prior knowledge, and the parameter bounds were chosen more narrowly, better results would, of course, be expected. For a fair comparison, however, the initial arrangement must be chosen without such additional information. The reason why the optimization was not able to reach the result reported in the main text, which was approximately twice as large, might be the large parameter space.

\subsection{Gradient-based optimization with a full-wave solver}
In the following, we discuss the optimization iterations performed with two commercial full-wave solvers. We selected solvers that can compute derivatives efficiently. Primarily, we are interested in the cost of a single iteration, i.e., computation of the value and derivatives. In this case, the second example from the main text, the infinitely periodic metasurface, is chosen for comparison. Because the initial spheres are small, requiring a very fine spatial discretization in the full-wave simulations and yielding very small objective values unsuitable for comparison, we set the sphere radius to 80~nm. We reproduce the case of a normally incident plane wave with a wavelength of 1050~nm, and keep all other parameters the same as in the main text.

\textbf{Finite Difference Time Domain solver}
We consider a full-wave solver Tidy3D~\cite{Liu:25}, a finite-difference time-domain electromagnetic solver with a Python interface and support for gradient evaluation. In this case, automatic differentiation is used only for Python-side operations, such as geometry parametrization. The gradients of the full-wave solve are obtained via the adjoint method, i.e., by performing an additional adjoint simulation. 

\begin{figure}[!htb]
  \centering
  \includegraphics[width=\linewidth]{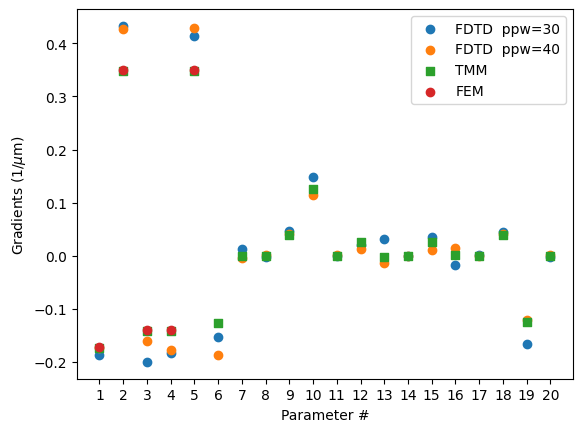}
  \caption{
Comparison of the gradients of the objective function, defined as the reflectance difference for orthogonal polarizations, with respect to the sphere radii and positions, for the T-matrix method (TMM, green square) and FDTD. FDTD results computed at two spatial resolutions, corresponding to 30 (blue circle) and 40 points (orange circles) per wavelength (ppw). The first five parameters are the radii of the five spheres, and the following 15 parameters are the $x$-, $y$-, and $z$-coordinates of the spheres. Additionally, for the radii, we also show the gradients obtained with a FEM solver (red circles). 
}
\label{grads}\end{figure}

Since the sphere primitive is not differentiable with respect to the required parameters, a differentiable approximation of the spheres was used instead. Specifically, the sphere boundaries were represented by smooth $\tanh$ functions. Furthermore, we set the domain size to 2100~nm, and set the minimum number of grid points per wavelength to 30. Computations were also performed with an increased number of grid points, i.e., 40, but the computed gradients depend only marginally on the chosen spatial resolution. Periodic boundary conditions were imposed in the transverse directions, and perfectly matched layers were used along the propagation direction.  The objective value coincides at approximately 0.0021, however,  the reflectance values are larger than the T-matrix method results (0.034 and 0.036 for $x$- and $y$-polarized illumination, respectively), reaching 0.0384 and 0.040. Meanwhile, keeping the same setup but replacing approximate spheres with primitive spheres with a locally overriden mesh of 4~nm produces values matching the referential values obtained by the T-matrix method. The fact that such a grid density does not improve the results for the approximate spheres indicates that the limiting factor is the sphere approximation itself. Thus, the current implementation does not provide the best achievable performance of the solver.

The computational times for the approximate spheres setup in the FDTD are 64 s for the forward simulation and 293 s for the forward-and-backward evaluation, which is clearly larger than the per-evaluation cost reported for the T-matrix method used in the main text.  It should be noted that the Tidy3D computations were performed on remote GPU hardware provided by the solver, offering greater computation resources, except for the spheres setup stage, which was performed on the same machine as the previous benchmarks. Therefore, the corresponding runtimes are not directly comparable.

\textbf{Finite element method solver}
Next, we performed a single optimization iteration using a frequency-domain finite-element electromagnetic solver JCMsuite. JCMsuite provides built-in sensitivity analysis, allowing the computation of shape and material derivatives of the electromagnetic fields and derived quantities up to second order. However, obtaining the gradients with respect to the positions of the spheres was not possible. Nevertheless, we performed simulations for the same setup, using a computational domain size of 600x600x500~nm$^3$, and set the background mesh size to 40~nm, while the sphere mesh size is 15~nm at the boundaries, and 25~nm in the interior. Curvilinear elements were disabled, and a finite element degree of 2 was used. The total time required to compute only the objective value was 1104 seconds, while the full computation, including gradients, took 2422 seconds, substantially exceeding the semi-analytical computation time.

The gradients obtained with the three methods are shown in Fig.~\ref{grads}. The FEM results are in excellent agreement with the T-matrix results. The FDTD gradients are also close, although an offset relative to the other methods remains. As discussed before, since the FDTD approach uses a finite computational grid and a smoothed tanh-based representation of the sphere boundaries, while the T-matrix method considers ideal spheres with exact interfaces, an exact agreement was not expected. The inclusion of two FDTD resolutions shows that the gradients are weakly affected by further mesh refinement, whereas an offset relative to the T-matrix results remains.
\section{Experimental realization at microwave frequencies}
\label{sec:Experimental realization at microwave frequencies}
In this supplement, we describe efforts that demonstrate the functionality of the designed devices at microwave frequencies. This required us to perform initial simulations adjusted to the experimental constraints that are described first. Then, we outline details on how the samples were fabricated, and finally, we explain the details of the characterization. A summary of these results had been included in the main document.  

\subsection{Simulation}
Although the initial designs in the main part of the manuscript were described at optical frequencies, we have implemented, verified, and scaled an experimental setup that allows for observation at microwave frequencies. Thanks to the scale invariance of Maxwell's equations, this is generally a valid approach, as long as the material properties are identical. Since the material properties are different than those considered in the main text, we resimulated the electromagnetic response of the samples with the experimentally accessible material properties, and we describe these simulations in this subsection at first. 

First, the frequency around 3 GHz was taken as the design frequency, corresponding to a wavelength of \SI{10}{\centi\meter}. That is in agreement with the experimental equipment available to us. The spheres were made from Acrylonitrile
butadiene styrene (ABS), a common thermoplastic polymer. As retrieved from dedicated measurements described in the experimental subsection further below, the material was characterized by a permittivity of $\varepsilon=\varepsilon'+i\varepsilon''=2.54+i0.26$. The approach to optimize the spheres' positions and radii, as described in the main text, was applied again using this adjusted permittivity value. The main purpose of the design has been to optimize the forward-to-backward scattering ratio. Identical to the scenario considered in the main text, the cluster consisted of six spheres.  The incident wave is $y$-polarized as in the main text. The resulting positions and radii, normalized by wavelength, are shown in the tables below. For comparison, the table also displays the geometrical details of the initial cluster used for reference purposes. All geometrical parameters are normalized to a wavelength of \SI{10}{\centi\meter}.

As the integrated quantity employed during optimization was not measurable with our experimental setup, the subsequent analysis is restricted to scattering at 0° and 180° (forward and backward, respectively).

\begin{table}[h]
\centering
\caption{Positions and radii of the spheres that make up the clusters. The details of the initial cluster are presented, along with those of the optimized cluster. Normalization is done with respect to the design wavelength of \SI{10}{\centi\meter}.}
\label{tab:clusters}

\begin{subtable}{0.47\textwidth}
\centering
\caption{Initial cluster}
\label{tab:init}
\scriptsize
\begin{tabular}{@{}lrrrr@{}}
\toprule
Sphere & x & y & z & r \\
\midrule
\#1 & 0.50 &  0.000 &  0.00 & 0.10 \\
\#2 & 0.25 &  0.433 &  0.00 & 0.10 \\
\#3 &-0.25 &  0.433 &  0.00 & 0.10 \\
\#4 &-0.50 &  0.000 &  0.00 & 0.10 \\
\#5 &-0.25 & -0.433 &  0.00 & 0.10 \\
\#6 & 0.25 & -0.433 &  0.00 & 0.10 \\
\bottomrule
\end{tabular}
\end{subtable}
\hfill
\begin{subtable}{0.47\textwidth}
\centering
\caption{Optimized cluster}
\label{tab:opt}
\scriptsize
\begin{tabular}{@{}lrrrr@{}}
\toprule
Sphere & x' & y' & z' & r' \\
\midrule
\#1 & 0.305 & -0.000 &  0.110 & 0.130 \\
\#2 & 0.232 &  0.281 & -0.109 & 0.218 \\
\#3 &-0.234 &  0.296 & -0.105 & 0.218 \\
\#4 &-0.308 &  0.000 &  0.113 & 0.130 \\
\#5 &-0.237 & -0.280 &  0.107 & 0.219 \\
\#6 & 0.237 & -0.280 & -0.107 & 0.219 \\
\bottomrule
\end{tabular}
\end{subtable}

\end{table}

\begin{figure}[ht]
\centering
\begin{subfigure}{0.49\linewidth}
  \centering
\includegraphics[width=\linewidth]{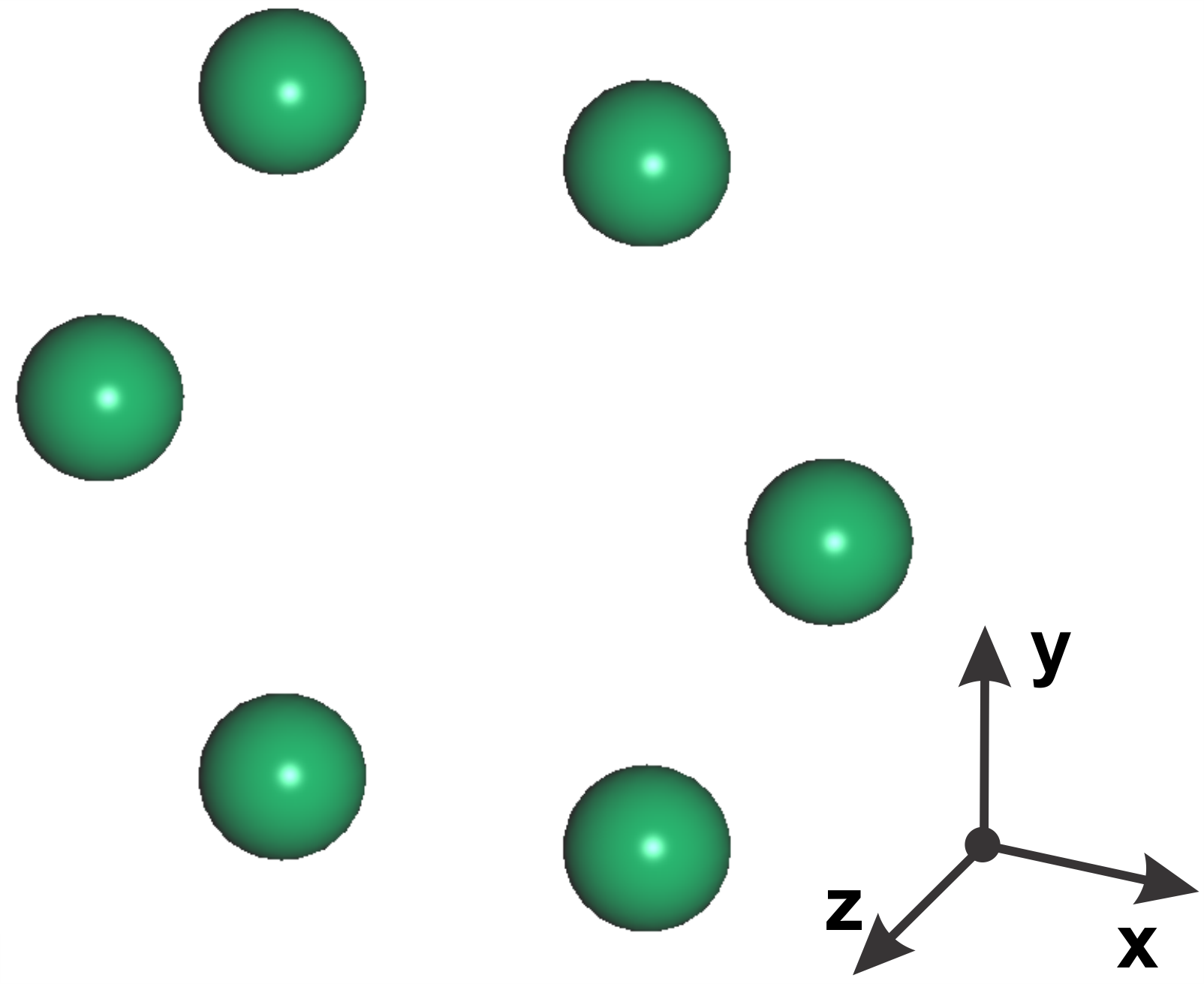} %
  \caption{}\label{fig:1a}
\end{subfigure}\hfill
\begin{subfigure}{0.49\linewidth}
  \centering
  \includegraphics[width=\linewidth]{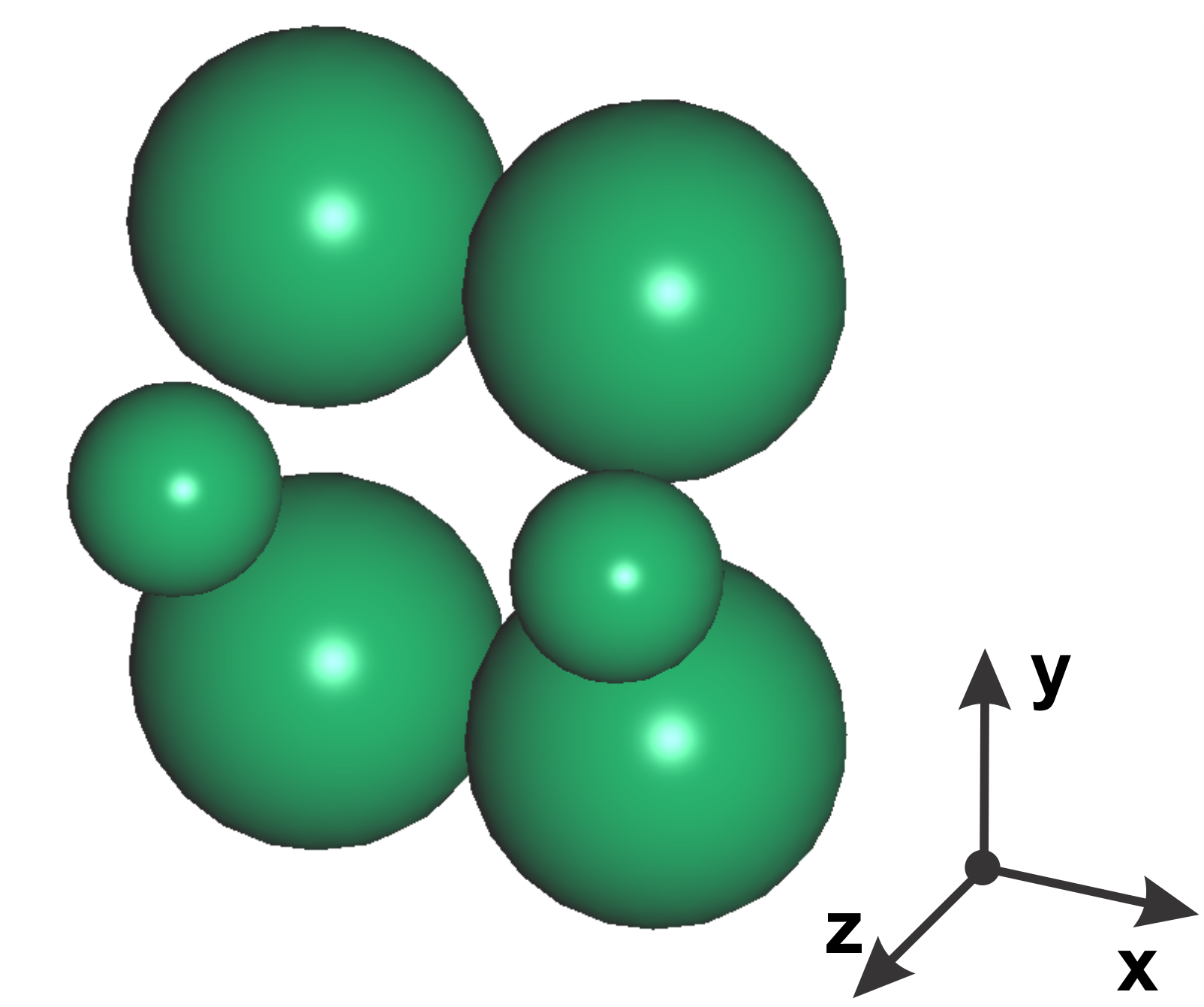}
  \caption{}\label{fig:1b}
\end{subfigure}

\vspace{2pt}

\begin{subfigure}{0.49\linewidth}
  \centering
  \includegraphics[width=\linewidth]{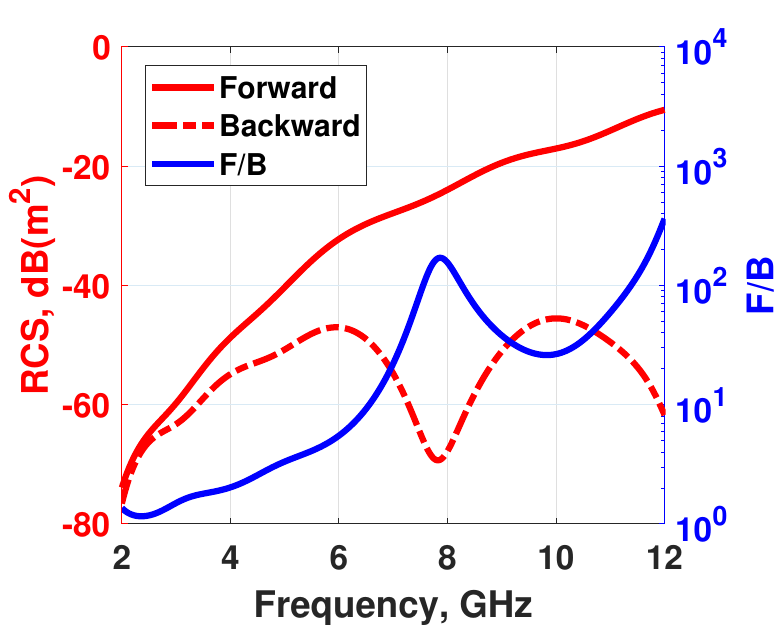}
  \caption{}\label{fig:1c}
\end{subfigure}\hfill
\begin{subfigure}{0.49\linewidth}
  \centering
  \includegraphics[width=\linewidth]{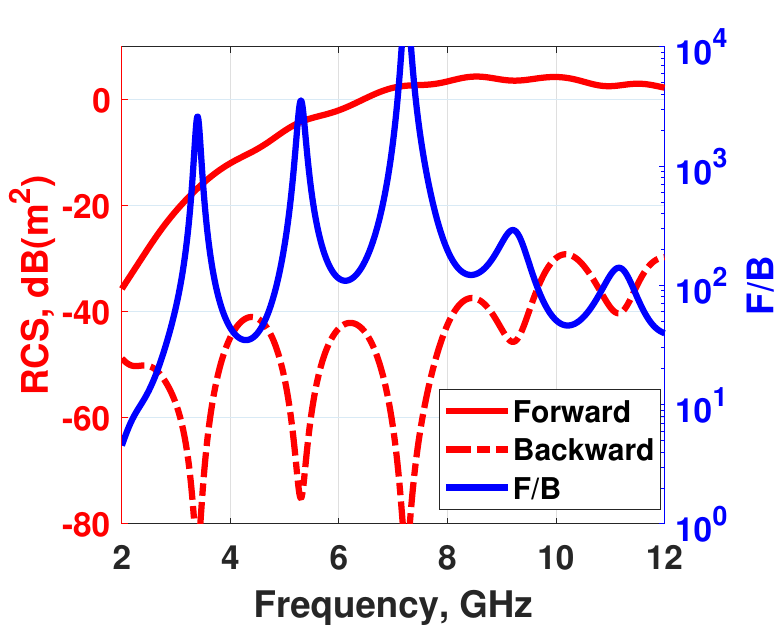}
  \caption{}\label{fig:1d}
\end{subfigure}

\vspace{2pt}

\begin{subfigure}{0.49\linewidth}
  \centering
  \includegraphics[width=\linewidth]{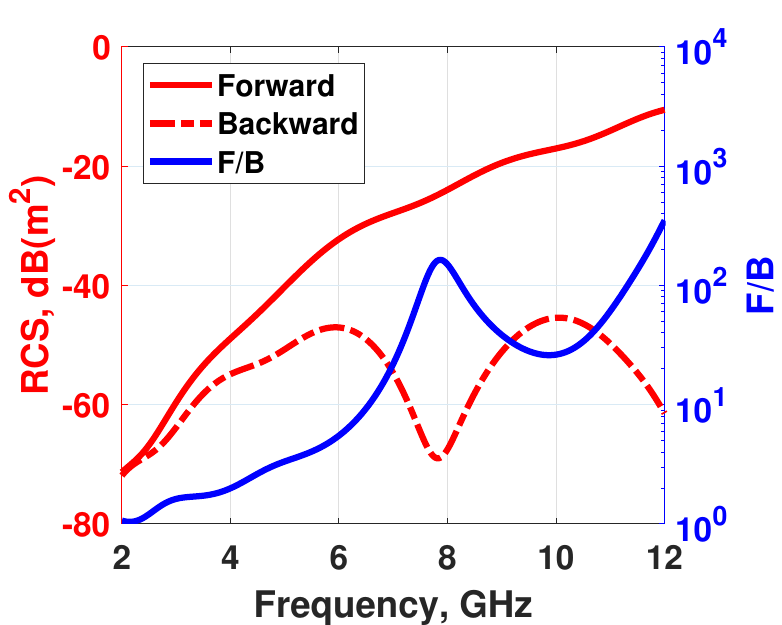}
  \caption{}\label{fig:1e}
\end{subfigure}\hfill
\begin{subfigure}{0.49\linewidth}
  \centering
  \includegraphics[width=\linewidth]{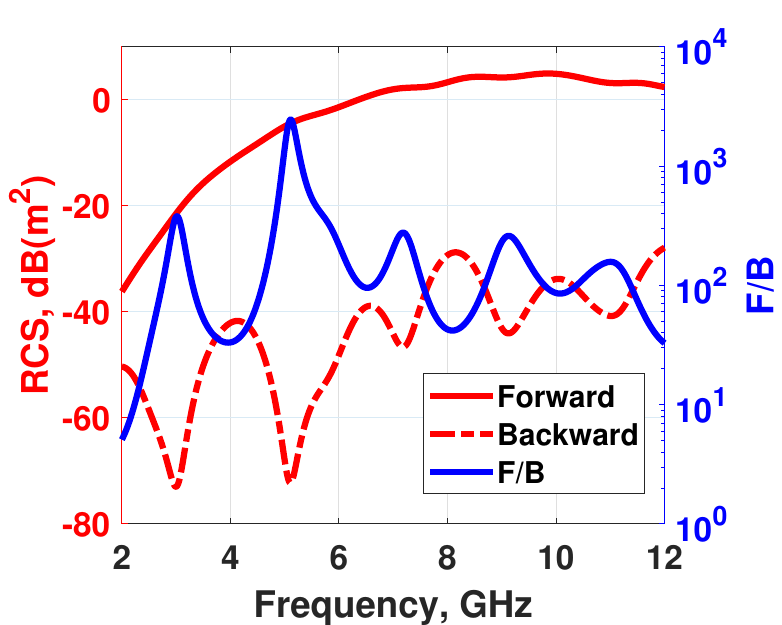}
  \caption{}\label{fig:1f}
\end{subfigure}

\caption{CST Microwave Studio models of (a) the initial and (b) the optimized cluster, plane wave propagation is along z-axis. (c)-(d) RCS spectra into the forward and backward direction and the forward-to-backward (F/B) ratio for an $x$-polarization of the illuminating plane wave. (e)-(f) Same plots, but for a $y$-polarization of the illuminating plane wave. Please be reminded that the sample was optimized for an operation at 3 GHz.}
\label{fig:fig1}
\end{figure}

The initial and optimized configurations, as shown in Figs.~\ref{fig:fig1}\subref{fig:1a}-\subref{fig:1b},  were also simulated in CST Microwave Studio. The change in the computational suite was done because it has capabilities aligned with our needs to compare the experimental results to numerical predictions. In the simulations, the structures are illuminated by a linearly polarized plane wave propagating along the z-axis. Both $x$- and $y$-polarization are considered separately. Two far-field radar cross-section (RCS) EM probes (for the forward and backward scattering) from the CST tool library were used to assess the directional scattering. The simulations were performed for the frequency range \SIrange{2}{12}{\giga\hertz} (the range corresponds to the capabilities of the laboratory equipment, and includes  \SI{3}{\giga\hertz} -- the design frequency). The RCS in dB(m\textsuperscript{2}) dispersions are shown in Figs.~\ref{fig:fig1}\subref{fig:1c}-\subref{fig:1f} for the initial and optimized clusters. Please focus on the design frequency of 3 GHz. As can be seen, the backward scattering is significantly lower than the forward scattering for both structures, while the optimized structure demonstrates an improvement of 10-40 dB(m\textsuperscript{2}) in forward scattering. 

\begin{figure}[h!]
\centering

\begin{subfigure}{0.48\linewidth}
  \centering
  \includegraphics[width=\linewidth]{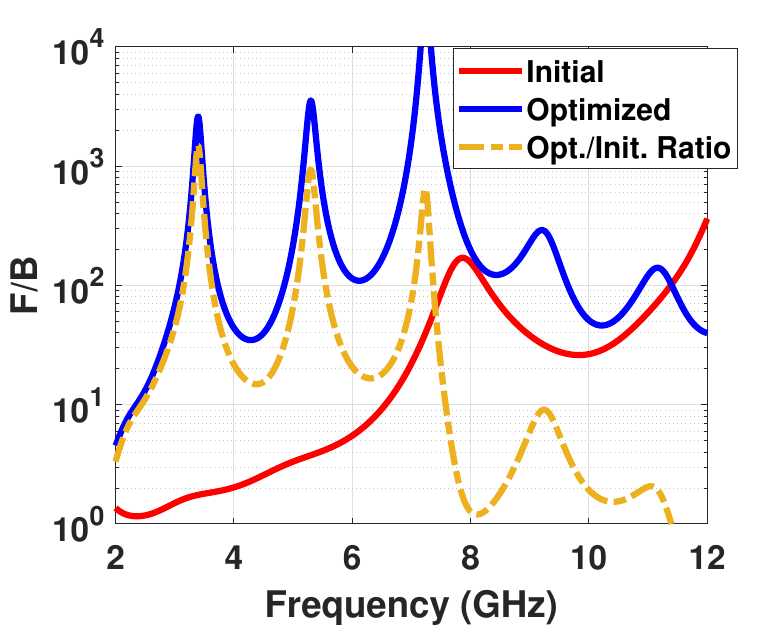}
  \caption{}\label{fig:2a}
\end{subfigure}\hfill
\begin{subfigure}{0.48\linewidth}
  \centering
  \includegraphics[width=\linewidth]{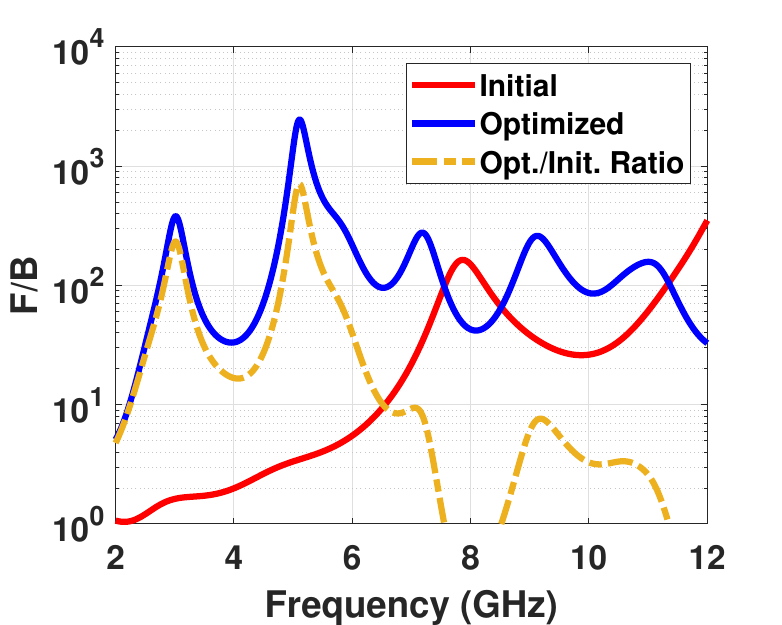}
  \caption{}\label{fig:2b}
\end{subfigure}

\vspace{3pt}

\begin{subfigure}{0.48\linewidth}
  \centering
  \includegraphics[width=\linewidth]{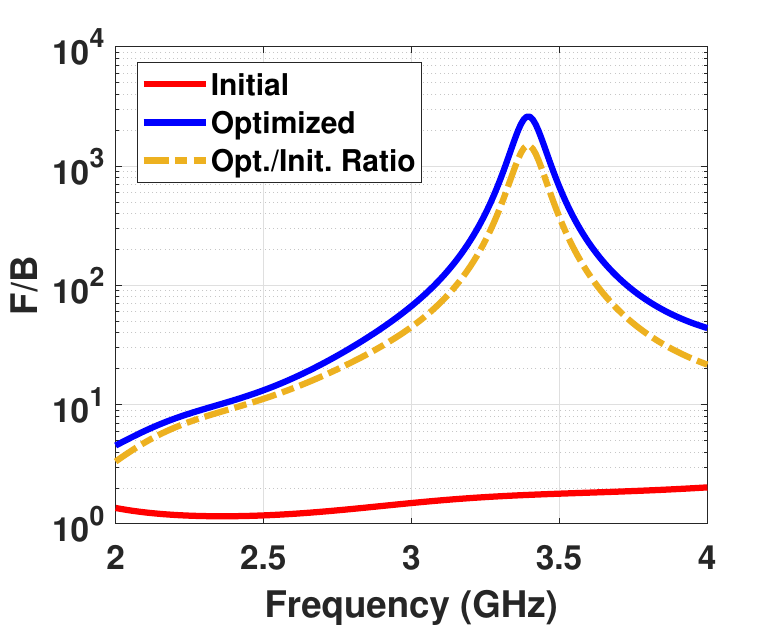}
  \caption{}\label{fig:2c}
\end{subfigure}\hfill
\begin{subfigure}{0.48\linewidth}
  \centering
  \includegraphics[width=\linewidth]{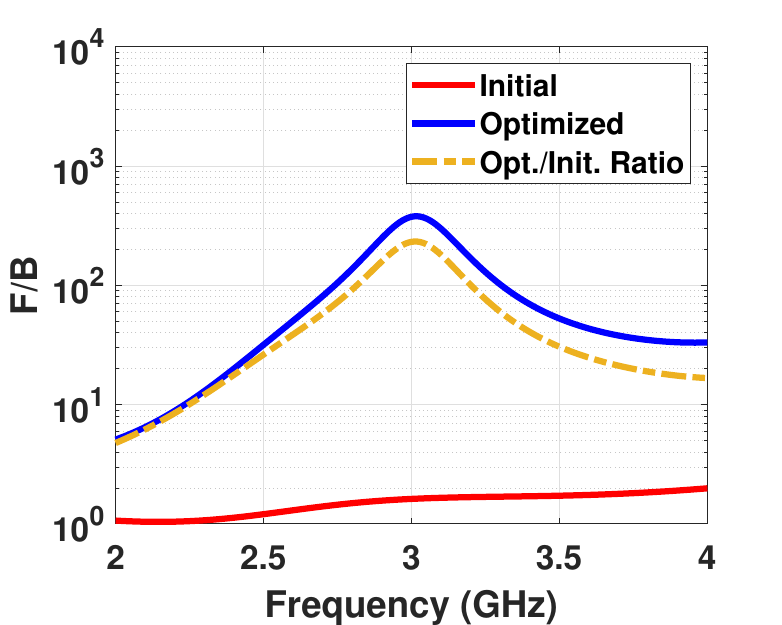}
  \caption{}\label{fig:2d}
\end{subfigure}

\caption{Simulated forward-to-backward spectra for the models of both structures (initial — solid blue, optimized — solid red) and forward-to-backward dispersion advantage of the optimized structure (dash-dotted) for an illumination with a plane wave propagating in z-direction and a polarization in (a) in x- and in (b) in y-direction, respectively. Zoomed-in plots around the frequency considered in the optimization are shown in (c) and (d), respectively.}
\label{fig:fb_disp}
\end{figure}

Figure~\ref{fig:fb_disp} demonstrates the forward-to-backward ratios for each structure (initial -- solid blue and optimized -- solid red) and their ratio to estimate the advantage of the optimized structure (dash-dotted plot). For the case of $x$-polarization, an improvement of 1486 times is obtained at \SI{3.39}{\giga\hertz} (Fig.~\ref{fig:fb_disp}\subref{fig:2a}, \subref{fig:2c}), while for the $y$-polarization it is 233 times at  \SI{3.}{\giga\hertz} (Fig.~\ref{fig:fb_disp}\subref{fig:2b}, \subref{fig:2d}). This result is achieved because of the appearance of a significant maximum in the forward-to-backward ratio for the optimized structure around the design frequency.




\subsection{Experiments }
\subsubsection*{Samples fabrication}
The experimental samples were designed in CST Microwave Studio in accordance with the sizes scaled to the microwaves and exported in STL format for subsequent 3D-printing with BCN3D Sigma 3D Printer. ABS was used as the material of the spheres. 

\begin{figure}[h]
  \centering
  \begin{subfigure}{0.48\linewidth}
    \centering
    \includegraphics[width=\linewidth]{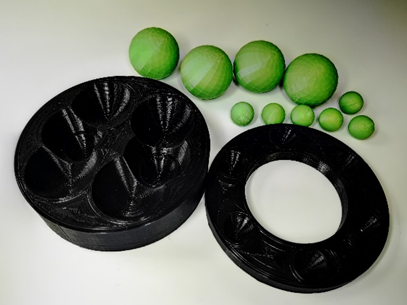}
    \caption{}
    \label{fig:3a}
  \end{subfigure}
  \hfill
  \begin{subfigure}{0.48\linewidth}
    \centering
    \includegraphics[width=\linewidth]{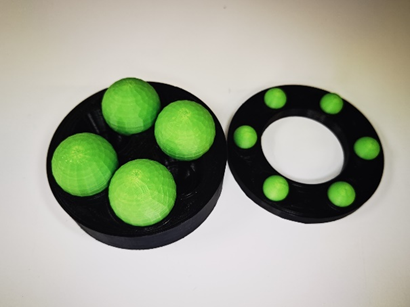}
    \caption{}
    \label{fig:3b}
  \end{subfigure}
  \caption{(a) A matrix for spherical electromagnetic scatterers made of black PLA plastic; and (b) an example of the spheres' arrangement within the system before foam encapsulation.}
  \label{fig:fab}
\end{figure}

The electromagnetic properties of ABS were tested using the open-end coaxial probe method. The method allows us to estimate the complex permittivity. Measurements were made using the Agilent 85070E dielectric probe kit and an E8361C 10 MHz–67 GHz PNA Network Analyzer. Measuring the complex permittivity is crucial for accurately reproducing experimentally measured quantities in simulations. 
Precise positioning of spatially separated spherical scatterers and their vertical arrangement parallel to the plane of a vertically polarized incident plane wave was achieved by utilizing a special plastic matrix. This matrix held the spheres in place before encapsulation (or casting) with a foam material (Fig.~\ref{fig:fab}). 

A complementary model of the initial scatterer group matrix was developed in CST, which assumed the placement of the spheres in ``niches'' or pockets, ensuring a precise fixed distance between them. After 3D printing of the original spheres from ABS plastic and the matrix model from PLA plastic (any other 3D printing filament with sufficient rigidity is also suitable for the matrix), all spheres were inserted into the matrix at their respective positions. The matrix was then filled with a foam material composed of the following components: Methylene diphenyl diisocyanate, Polyether polyol, and 1,1-Dichloro-1-fluoroethane (HCFC-141b). The dielectric parameters of the resulting foam are very close to those of air, as the solidified foam is predominantly composed of air (due to its porous structure).

\begin{figure}[tbh]

\begin{subfigure}{0.2\textwidth}
  \centering\includegraphics[width=\linewidth]{figs/Experiment/initial_cluster_experiment.jpeg}
  \caption{}\label{fig:4a}
\end{subfigure} \hspace{0.1\linewidth} %
\begin{subfigure}{0.2\textwidth}
  \centering\includegraphics[width=\linewidth]{figs/Experiment/final_cluster_experiment.jpeg}
  \caption{}\label{fig:4b}
\end{subfigure}


\begin{subfigure}{0.44\textwidth}
  \centering\includegraphics[width=\linewidth]{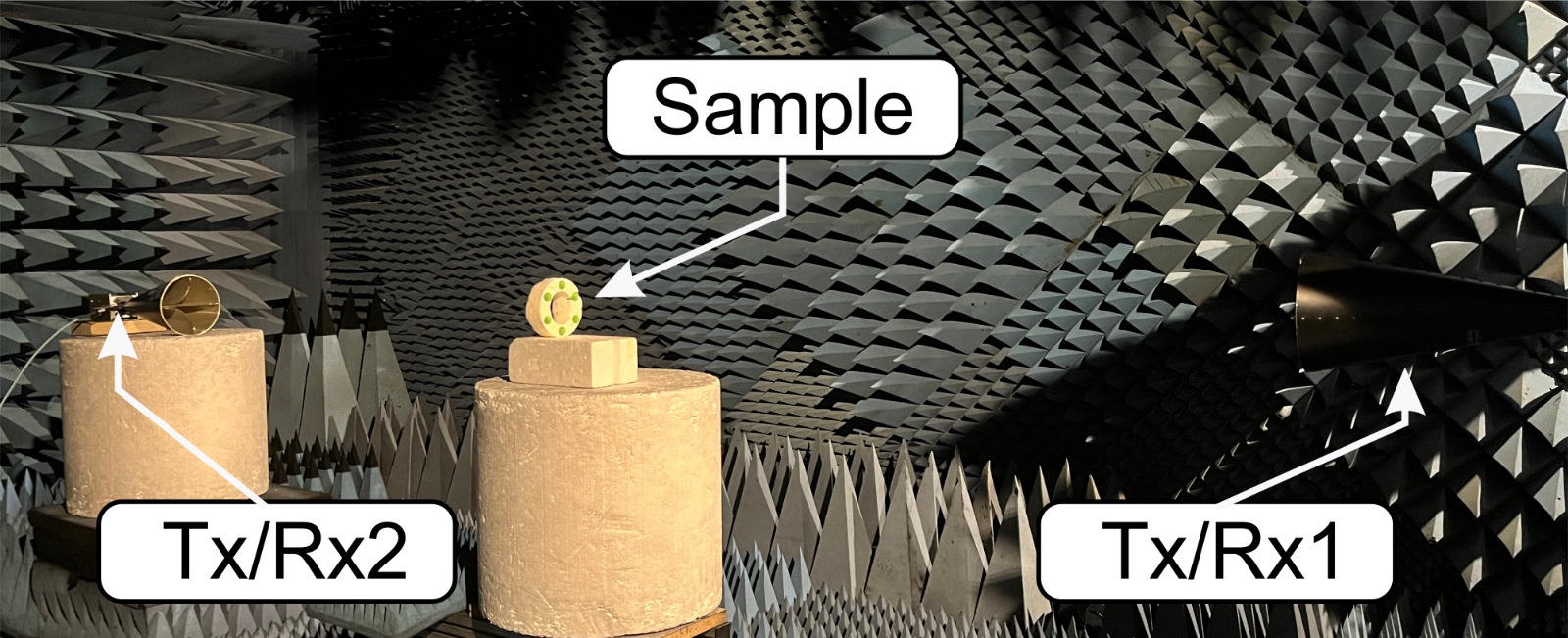}
   \caption{}\label{fig:4c}
\end{subfigure}
\caption{Fabricated samples using 3D-printing and foam hosting: (a) the initial cluster and (b) optimized cluster. (c) The experimental setup in the anechoic chamber with two horn antennas for testing the structures. }
\label{fig:7first}
\end{figure}

After filling the spherical scatterer matrix with the foam and its subsequent curing for 24 hours, the original matrix was removed from the scatterers, which became solidified and precisely fixed within the foam.
As a result of fabrication, the initial and optimized structures were prepared as shown in Figs.~\ref{fig:7first}\subref{fig:4a}-\subref{fig:4b}.

\subsubsection*{Samples characterization}

\begin{figure}[!ht]

\begin{subfigure}{0.22\textwidth}
  \centering\includegraphics[width=\linewidth]{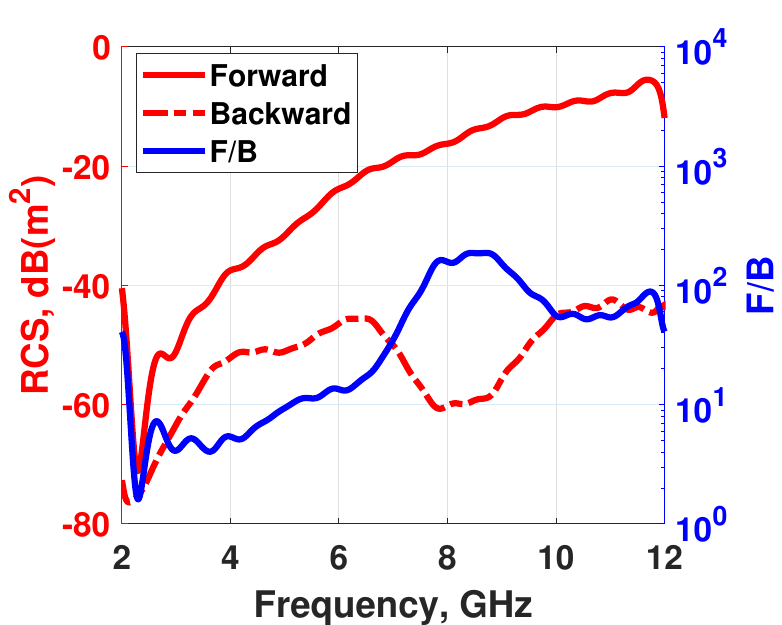}
  \caption{}\label{fig:4d}
\end{subfigure} \hspace{0.05\linewidth}
\begin{subfigure}{0.22\textwidth}
  \centering\includegraphics[width=\linewidth]{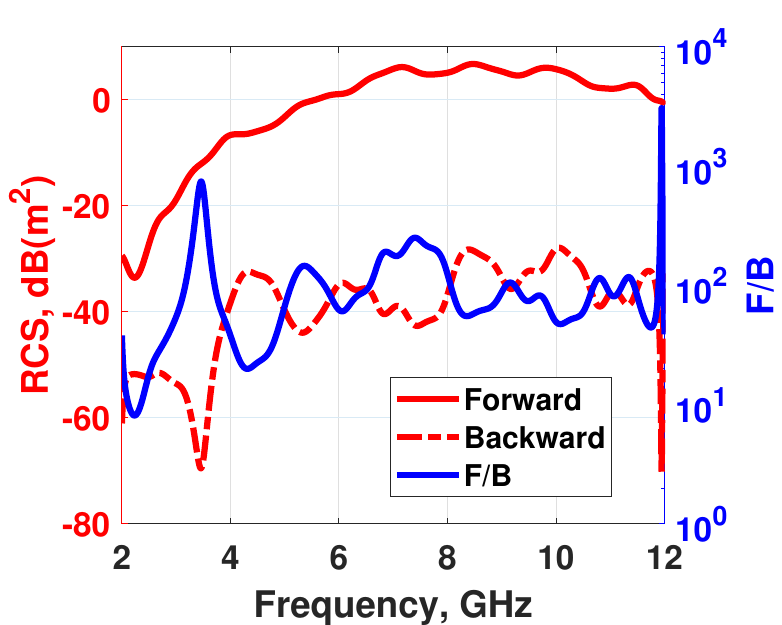}
 \caption{}\label{fig:4e}
\end{subfigure}

\vspace{0.6em}

\begin{subfigure}{0.22\textwidth}
  \centering\includegraphics[width=\linewidth]{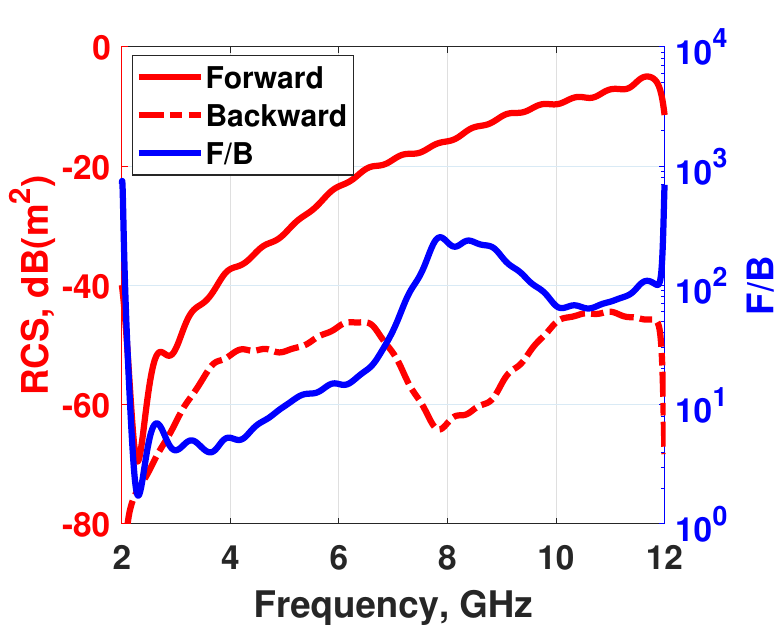}
\caption{}\label{fig:4f}
\end{subfigure} \hspace{0.05\linewidth}
\begin{subfigure}{0.22\textwidth}
  \centering\includegraphics[width=\linewidth]{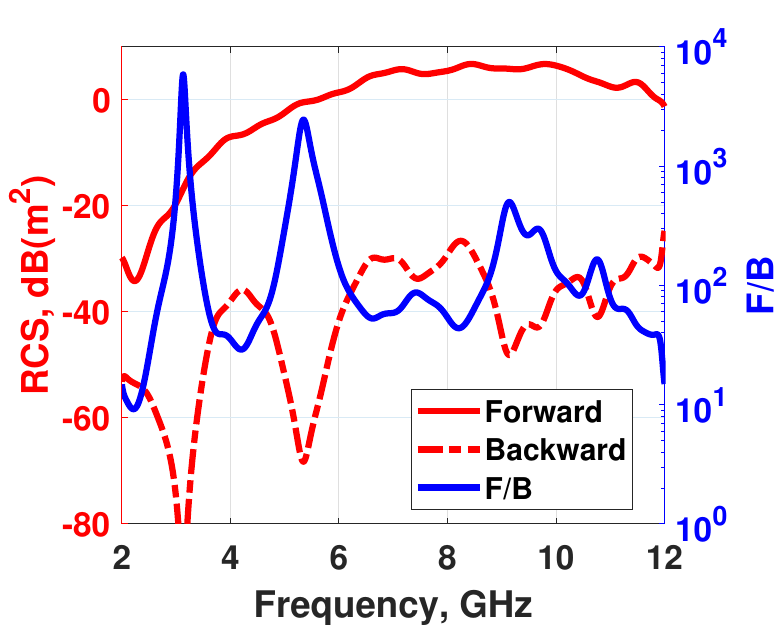}
  \caption{}\label{fig:4g}
\end{subfigure}

\caption{ (a)-(b) Experimentally obtained RCS spectra for the initial and optimized structures, respectively, including forward and backward scattering and their F/B ratio for the $x$-polarization; (c)-(d) the same plots but for the $y$-polarization.}
\label{fig:7second}
\end{figure}

The experimental setup to characterize the samples is assembled in a certified anechoic chamber. It consists of two horn antennas (NATO IDPH-2018, \SIrange{2}{18}{\giga\hertz}) spaced \qty{3}{\metre} apart, connected to a Keysight P9374A Network Analyzer (300 kHz–20 GHz) that can record S-matrix values in the complex representation for $x$- and $y$-polarizations, see Fig.~\ref{fig:7first}\subref{fig:4c}. Each measurement used network analyser settings of 5 dBm output power, \SI{100}{\kilo\hertz} input bandwidth, and 16001 points. Each sample was installed exactly at the midpoint between the antennas on a Styrofoam stage, which is transparent to microwave propagation. For quantitative measurements, a brass calibration disk of \SI{10}{\centi\meter} diameter has been used. Time gating post-processing was applied to reduce the multipath impact further. 

\begin{figure}[!ht]
\centering
\begin{subfigure}{0.49\linewidth}\centering
  \includegraphics[width=\linewidth]{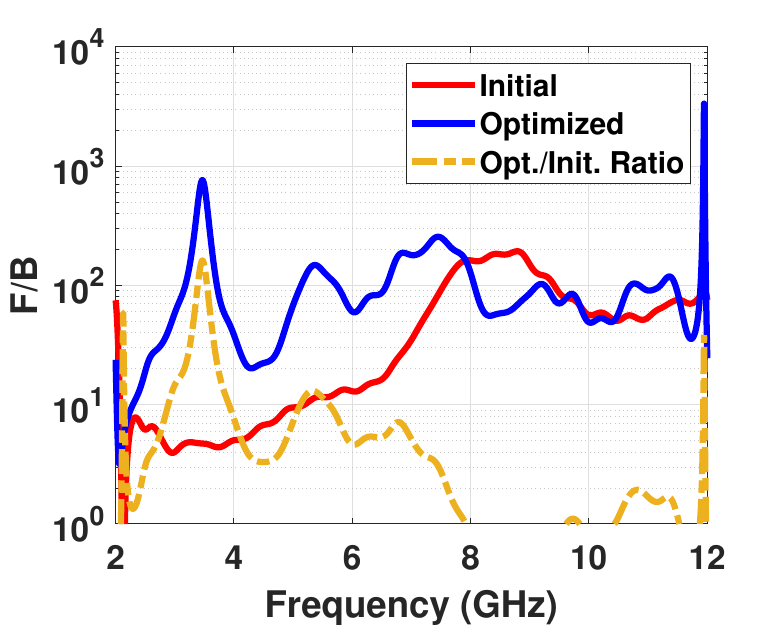}\caption{}\label{fig:5a}
\end{subfigure}\hfill
\begin{subfigure}{0.49\linewidth}\centering
  \includegraphics[width=\linewidth]{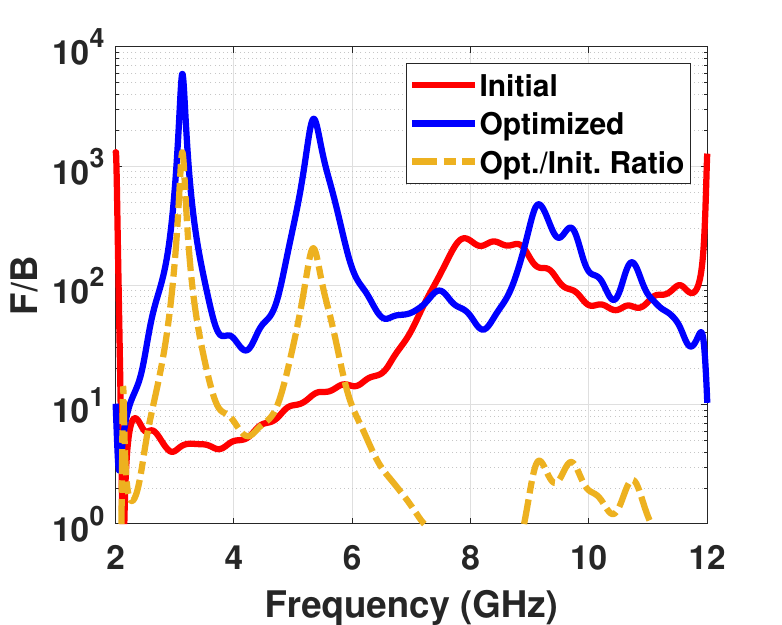}\caption{}\label{fig:5b}
\end{subfigure}
\vspace{2pt}
\begin{subfigure}{0.49\linewidth}\centering
  \includegraphics[width=\linewidth]{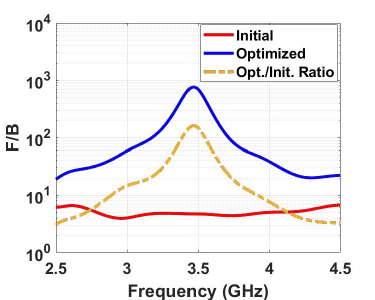}\caption{}\label{fig:5c}
\end{subfigure}\hfill
\begin{subfigure}{0.49\linewidth}\centering
  \includegraphics[width=\linewidth]{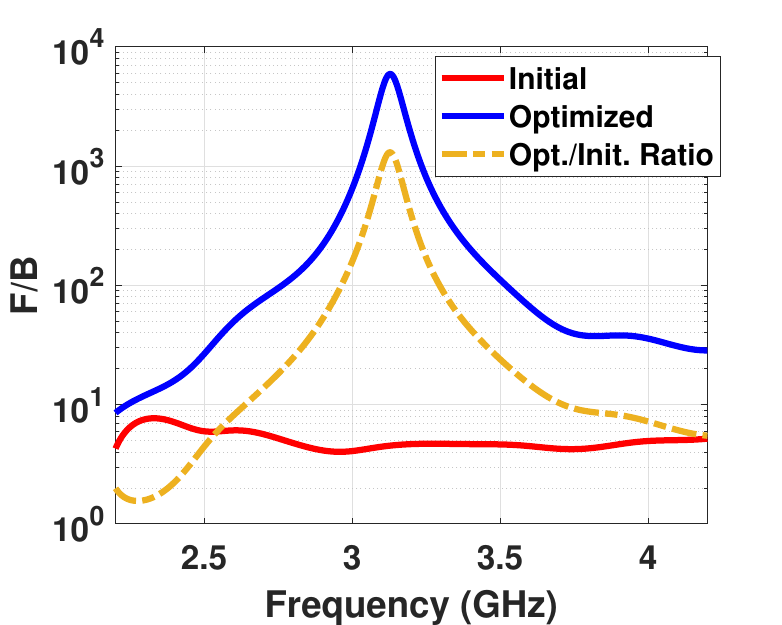}\caption{}\label{fig:5d}
\end{subfigure}
\caption{Forward-to-backward (F/B) scattering for both experimental structures—initial (solid blue) and optimized (solid red)—and the forward-to-backward scattering ratio of the optimized to the initial cluster (dash-dotted), respectively, for the (a) $x$- and (b) $y$-polarization. A zoomed-in version of the same figure around the optimization frequency in (c) and (d), respectively.}
\label{fig:2x2}
\end{figure}

From the calculated RCS spectra of forward and backward scattering, along with their F/B ratio, shown in Fig.~\ref{fig:7second}\subref{fig:4d}-\subref{fig:4g}, one can see a good agreement between the results and the simulations. 
As in the simulations, a strong minimum is observed in the backward scattering spectra of the optimized structure near the design frequency (Figs.~\ref{fig:7second}\subref{fig:4e}, and \subref{fig:4g}). Therefore, the F/B ratios for the optimized structure are characterized by maxima (Fig.~\ref{fig:2x2} -- the blue plots) compared with the initial structure (Fig.~\ref{fig:2x2} -- the red plots). In the experimental study, an enhancement of\textasciitilde160 was achieved for the $x$-polarization (Fig.~\ref{fig:2x2}\subref{fig:5c}) at the frequency \SI{3.47}{\giga\hertz},  and \textasciitilde1300 for the $y$-polarization at the frequency \SI{3.13}{\giga\hertz} (Fig.~\ref{fig:2x2}\subref{fig:5d}), compared to the simulation analysis in Fig.~\ref{fig:fb_disp}. We note that the magnitude of the ratio is essentially determined by how small the detected backward-scattering signal is at this frequency. Therefore, deviations may arise not only from differences between the fabricated and simulated structures, but also from the limited signal level in the measurement.

Taking all these results, the experiments show that the methodology described in the main text can be used to optimize structures whose functionality is verifiable on experimental grounds. The agreement between measurements and simulations reveals, moreover, the functionality of the design.

\subsection*{Sensitivity analysis}

\begin{figure}[!htbp]
\centering
  \includegraphics[width=\linewidth]{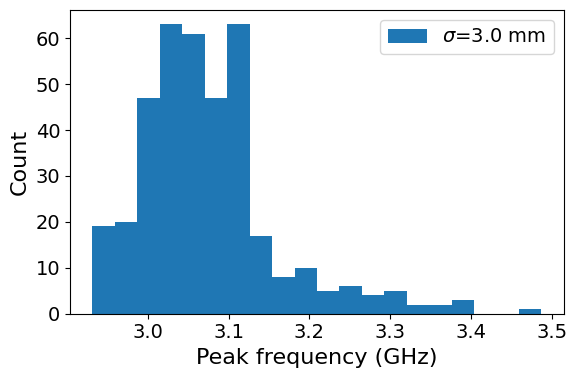}\caption{Distribution of peak frequencies for $y$-polarized illumination obtained from random perturbations of the sphere positions with a standard deviation of 3~mm. Only non-overlapping arrangements were considered.
  }\label{fig:shift}
\end{figure}

We investigate here the impact of positional inaccuracies of the spheres on the primary experimental results, i.e., the spectral position of the peak frequency for the forward-to-backward scattering ratio. To this end, we perturb the sphere positions by displacements drawn from a uniform distribution with a standard deviation of 3~mm. The optimized sample from the main manuscript illuminated with a $y$-polarized light is considered. 
The resulting distribution in Fig.~\ref{fig:shift} suggests that positional inaccuracies of this magnitude can lead to peak frequencies close to the experimentally observed value. 
We find that uncertainty in the placement of the spheres leads to a modest shift of the resonance peak toward slightly higher frequencies compared to the designed value. This behavior is fully consistent with our experimental observations, where the measured resonances also appear at somewhat higher frequencies than those predicted by the initial simulations. Taken together, these results demonstrate that even small deviations in the spatial positioning of the spheres are very likely responsible for the observed discrepancy between the experimental and simulated resonance frequencies.

\end{document}